\documentclass[11pt]{article}

\usepackage{amsmath,amsfonts,bm,amssymb}
\usepackage[numbers]{natbib}
\usepackage{datetime,vector}
\settimeformat{ampmtime}
\usepackage[]{graphicx,hyperref}
\usepackage{subfigure}
\usepackage{color}

\newcommand{\C}{\mathbb{C}}

\renewcommand{\P}[1]{\operatorname{P}\left\{#1\right\}}
\newcommand{\Pn}{\operatorname{P}}
\newcommand{\E}[1]{\operatorname{E}\left[#1\right]}

\newcommand{\bG}{\bm{G}}

\newcommand{\vect}{\vec}

\newcommand{\bPhi}{\mathbf{\Phi}}
\newcommand{\bK}{\mathbf{K}}

\newcommand{\cms}{random backpropagations}

\newcommand{\herm}{^H}

\def \endprf{\hfill {\vrule height6pt width6pt depth0pt}\medskip}

\begin{document}

	\title{Compressive Matched-Field Processing}%
	\author{William Mantzel, Justin Romberg, Karim Sabra
	\thanks{ [willem, jrom, ksabra3]@gatech.edu}
	}
	\date{August, 2011}%
	\maketitle

\begin{abstract}
	
Source localization by matched-field processing (MFP) generally involves solving a number of computationally intensive partial differential equations.  
This paper introduces a technique that mitigates this computational workload by ``compressing'' these computations.  Drawing on key concepts from the recently developed field of compressed sensing, it shows how a low-dimensional proxy for the Green's function can be constructed by backpropagating a small set of random receiver vectors. Then, the source can be located by performing a number of ``short'' correlations between this proxy and the projection of the recorded acoustic data in the compressed space.
Numerical experiments in a Pekeris ocean waveguide are presented which demonstrate that this compressed version of MFP is as effective as traditional MFP even when the compression is significant. The results are particularly promising in the broadband regime where using as few as two random backpropagations per frequency performs almost as well as the traditional broadband MFP, but with the added benefit of generic applicability. That is, the computationally intensive backpropagations may be computed offline independently from the received signals, and may be reused to locate any source within the search grid area.

\end{abstract}


\section{Introduction}
\label{sec:intro}
\subsection{Background and Motivation}
Matched field processing (MFP) continues to serve as one of the most widely used methods for localizing undersea targets acoustically. However, as the models governing undersea acoustic interactions become more sophisticated, often requiring fine-grain solutions to more complex partial differential equations, the tradeoff between run time and performance begins to worsen, perhaps unnecessarily. We will begin by discussing why this is the case and giving an overview of our approach to mitigate the problem.

MFP generalizes standard array beamforming methods (e.g. plane wave beamforming) for locating an acoustic source in a complex environment (such as a multipath shallow water waveguide). MFP has been studied extensively both theoretically and experimentally as described in several review articles \cite{tolstoy2000applications,baggeroer1993overview,baggeroer1988matched}. MFP is usually implemented by systematically placing a test point source at each point of a spatial search grid of $L$ candidate locations, computing the acoustic field (replicas) at all the elements of the receiver array and then correlating this modeled field with the data from the real point source whose localization is unknown to determine the best-fit location (see Fig.~\ref{fig:LaMer}). This approach works well when that the computational replica environment is sufficiently accurate. However, this direct implementation of MFP using brute force search would require $L$ computation runs which can become numerically cumbersome for large search space especially when simulating complex propagation environments. 

One alternative to this direct implementation of MFP is to use a ``backpropagation'' algorithm (also referred to as ``time-reversal imaging'') to locate the unknown source. In this case, a time-reversed version of the recorded data is used as an initial waveform excitation along the array aperture using the principle of superposition, and then subsequently ``backpropagated'' numerically in the replica environment towards the grid search area \cite{gruber2004time}. The unknown source location is then estimated from the maximum of the distribution of the backpropagated peak amplitude (or energy) across the grid search. Consequently, when compared to the direct implementation first mentioned, this backpropagation approach appears attractive at first glance, since it  requires one computational run per unknown source.   Nevertheless, this backpropagation approach becomes computationally expensive if multiple sources need to be located repetitively over the \textit{same} search grid as the number of required computational runs would grow proportionally. 
For instance, this may occur when one tries to locate a source moving along a long track throughout the search space. Indeed, in order to be able locate any source throughout the search space using $N$ receivers, MFP would require computing $N$ backpropagations by using sequentially each individual receiver as a backpropagation source \cite{tolstoy2000applications,baggeroer1993overview}. This would allow determining the full set of Green's functions associated with the channel between each search location and each receiver element. Alternatively, one could weight spatially the amplitude of the backpropagated signals along the receiver array using $N$ different orthogonal codes (e.g. obtained from an Hadamard basis).

This article develops instead a compressive MFP formulation which reduces this computational burden by pre-computing the backpropagation of a number $M\ll N$ of random test signals.  
The results of these backpropagations effectively encode the Green's function associated with the channel, and they can be re-used in subsequent localizations without any additional computational cost.
This approach is inspired by recent work in the field of compressed sensing \cite{candes06st,candes06ne,donoho2006compressed}, whose central message is that random projections provide an effective encoding for sparse signals.  The motivation for compressed sensing is typically concerned with reducing the cost of acquiring signals by shifting the workload from sensor hardware to software \cite{candes08in,romberg08im,healy08co}, and is natural in applications where physical measurements are expensive compared to numerical computations.
Here we explore a variation on this theme: mitigating the computational workload in software instead of the sensing workload in hardware.
The proposed compressive MFP allows us to estimate the underlying ambiguity function central to conventional MFP algorithms over the entire search space using only $M$ computational runs instead of $N$, an effective speedup of a factor of $N/M$. In practice, these $M$ simulations can be independently computed as a background process offline before the actual source signal is received. 

\vspace{2mm}
\noindent
\subsection{Related Work}
In this paper, we effectively demonstrate how classical localization procedures under a least-squares framework such as matched-field processing (MFP) may be solved in a reduced-dimensional space even without {\em a-priori} knowledge of the ``best'' dimension-reducing transform.
This property has been shown in similar forms in the mainstream canon of Compressed Sensing (CS) literature.
Davenport et al. have described a number of useful variations on the theme of CS \cite{davenport2010signal} including a matched filtering detector. They have also described the ``smashed filter'' that is designed primarily for classification between a finite number of sets, but could easily be extended to parametric estimation.
Wakin has also established some rigorous results on parameter estimation that relate the recovery properties of a general compressive estimation problem to the properties of the manifold that these parameters induce. This work could be used to analyze this problem via its manifold parameters \cite{wakin2010manifold}.

Carin et al. have utilized CS principles to show how a Green's function of a scattering field that is compressible in the wavelet domain may be recovered from a small set of measurements, though they use incoherence in their structured measurements to recover a scattering field, while we primarily care about the location of the source \cite{carin2008situ}. Likewise, Marengo et al. have applied compressed measurements to the scattering problem, utilizing the target-sparse model to improve their performance \cite{marengo2008compressive}.

Our work may also be viewed in the context of randomized SVDs \cite{halko2009finding}. In this field of research, the idea is to apply the matrix $\bm{A}$ to a series of random vectors $\Phi_m$ as $\bm{A} \Phi_m$ in order to determine the range space of $\bm{A}$. For example, Chaillat et al. show how the inverse medium problem can be simplified using a dimension reducing random projection and solving the inverse problem in the reduced range-space \cite{chaillat2010falms}. Similarly to this field, we apply the time-reversal or adjoint of the Green's function $\bG_{\omega}$ to random vectors in order to discover the range space of admissible ambiguity functions.

There is also a large amount of recent research performing multi-target tracking under the ``target-sparse'' assumption. That is, the methods propose to simultaneously localize several targets that lie on some grid (or generally some set of points) by solving an $\ell_1$ minimization program. The recovered support resulting from this optimization corresponds to the grid points that the various targets are estimated to occupy. All of this work dovetails in very nicely with the main results of Compressed Sensing, which can be effectively leveraged to prove that the targets may be perfectly localized with high probability. Often, the painstaking effort in these papers involves showing that the Restricted Isometry Property (RIP) holds for the observation matrix. For example, Fannjiang et al. show the conditions under which a sufficiently small coherence is achieved for perfect recovery \cite{fannjiang904compressed}. Gurbuz et al. show similar results for a {\em Compressive} beamformer, requiring a number of measurements on the order of the number of sources \cite{gurbuz-compressive}, but the application there is different in that they utilize a signal common to all sensors with an unknown time shift to localize their target in angle (assuming free space propagation), and apply the compression operator in time per-sensor instead of applying the operator across the range of sensors as we do. Also from a communications perspective, Cevher et al. demonstrate the relatively low amount of information to be transmitted for purposes of localization when using a Compressed Sensing framework \cite{cevher2008distributed}. These ``target-sparse'' approaches depend on targets lying exactly on the grid points. Also, by necessity these grid points must be spaced sufficiently far away from one another to avoid coherence-inducing correlations in the observation matrix. This creates a restrictive model of limited applicability. When a target is somewhere in between a set of grid points, the necessary conditions for recovery may not even {\em approximately} hold, similarly to how a discrete sinusoid corresponding to an off-grid point in the DFT will not be sparse in the frequency domain (or any other basis for any standard transform for that matter) due to DFT leakage. In contrast to this approach, we do not require our target to lie on a grid point. However, instead of promising perfect recovery, we instead content ourselves to claim that our target may be localized to within a small neighborhood of the actual source location, or at least the location found via deterministic means.

\subsection{Outline}

The remainder of the paper is organized as follows. Section~\ref{sec:MFP} briefly describes conventional MFP formulation for locating both single-frequency (narrowband) and broadband sources. Section~\ref{sec:cMFP} presents the corresponding compressive MFP (cMFP) formulation for both cases. 
Section~\ref{sec:results} presents numerical simulations in a Pekeris waveguide \cite[pg 540--552]{jensen1994computational} illustrating the performance of cMFP in comparison to the conventional MFP results including the effects of additive ambient noise to the data and model mismatch due to uncertain knowledge of the actual environment.  Section~\ref{sec:adaptive} extends this compressive approach to adaptive MFP.  Section~\ref{sec:conclusion} summarizes the findings and conclusions drawn from this study.


\section{Conventional MFP}
\label{sec:MFP}

\begin{figure*}
\centerline{ 
\includegraphics[width=120.0mm]{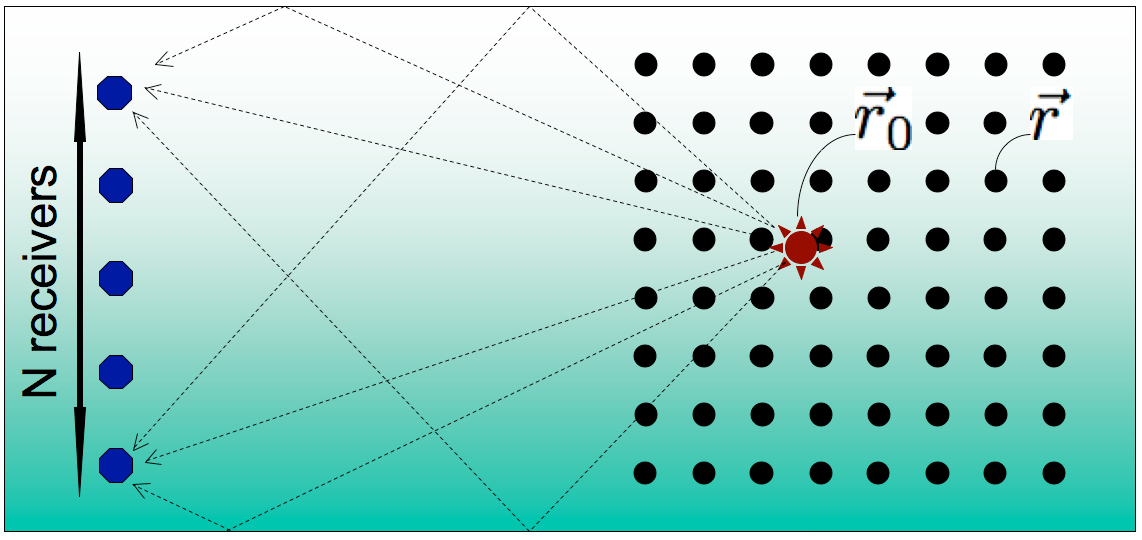}} \hspace{6mm}
\caption{\small\sl 
Schematic of  a matched-field processing implementation in an ocean waveguide. The signal transmitted by a source (star symbol) located at an unknown location $\vec r_0$ is recorded along a $N$ elements receiver array after multipath propagation. Using a computational model of the original ocean waveguide, the location $\vec r_0$ may be inferred  by matching the actual received signals with the simulated replica waveforms obtained from varying the test source location (dot symbols) $\vec r$ throughout the search grid area.}
\label{fig:LaMer}
\end{figure*}

A brief summary of the conventional MFP formulation is presented hereafter based on the standard solution of the linearized wave equation. The acoustic pressure field $y(\vec r,t)$  at a fixed point $\vec r$ and time $t$ produced by a point source  located at $\vec r_0$ satisfies:
\begin{equation}
	\frac{1}{c^2(\vec r)}\frac{\partial ^2 y(\vec r,t)}{\partial t^2} - \nabla^2 y(\vec r,t) =
	\alpha(t) \delta(\vec r -\vec r_0)
	\label{eq:Wave_Eqn}
\end{equation}
where $c(\vec r)$ is the speed of sound and $\alpha(t)$ is the signal emitted by the source. The time-domain Green's function for the same environment $ g(\vec r,\vec r_0,t)$ is, by definition, the solution of Eq. (\ref{eq:Wave_Eqn}) for a impulsive point source (i.e. for $\alpha(t)=\delta(t)$) that satisfies all boundary conditions \cite[pg 540--552]{jensen1994computational}. Using Eq. (\ref{eq:Wave_Eqn}) (and assuming that the radiation condition applies as $\|\vect{r}\| \rightarrow \infty$) the Fourier transform of the recorded pressure field at $\vec r_n$, the $n^{th}$ element of a receiver array ($n=1..N$) (see Fig. \ref{fig:LaMer}), is denoted $y_{\omega}(\vec r_n)$ and given by:
\begin{equation}
	y_\omega (\vec r_n)= \alpha_{\omega} g_\omega(\vec r_n,\vec r_0 )
	\label{eq:FOURIER_Sol}
\end{equation}
where $\omega$ is the frequency. The variables $\alpha_{\omega}$ and $g_\omega(\vec r_n,\vec r_0 )$ denote respectively the Fourier transform of the source signal and time-domain Green's function. Using vector notation,  Eq. \eqref{eq:FOURIER_Sol} can be restated as:
\begin{equation}
	Y_\omega = \alpha_{\omega} G_\omega(\vec {r_0}),
	\label{eq:themodel}
\end{equation}
where $Y_\omega$ is a ($N\times 1$) column vector obtained by stacking the complex amplitudes $y_\omega (\vec r_n)$ measured along the receiver array. Similarly, the ($N\times 1$) column vector $G_\omega(\vec {r})$ contains Green's functions  $g_\omega(\vec r_n,\vec r )$  between the  $N$ receiver array elements and a source located at $\vec {r_0}$. Note that the position vectors are written in lowercase letters with arrows and the column vectors are written with capital letters in the remainder of this article.  

\subsection{Single-Frequency MFP}

We start by considering the simplest MFP that works from measurements at a single frequency $\omega$ (as in \eqref{eq:themodel}), known as the harmonic (or narrowband) formulation.
Given a set of measurements $Y_{\omega} \in \C^N$ across the $N$ receivers at frequency $\omega$, we search for the location $\vec{r}$ in our region of interest $\mathcal{R}$ (and complex source amplitude $\beta$) that best accounts for these measurements by solving the least-squares problem
\begin{equation}
	\label{eq:ls-single}
	\arg\min_{\vec{r}\in\mathcal{R}}\min_{\beta\in\C}~\|Y_{\omega} - \beta G_{\omega}(\vec{r})\|^2.
\end{equation}
With the location $\vec{r}$ fixed, the inner optimization problem is simply finding the closest point on the line spanned by $G_{\omega}(\vec{r})$ to the point $Y_{\omega}$.  Plugging in the closed-form solution to this problem (see Appendix~\ref{app:closest-point}), the problem above reduces to: 
\begin{equation}
	\label{eq:amb-norm}
	\arg\min_{\vec{r}}\|Y_{\omega}\|^2 - \frac{|Y_{\omega}^H G_{\omega}(\vec{r})|^2}{\|G_{\omega}(\vec{r})\|^2} 
	~=~
	\arg\max_{\vec{r}}\frac{|Y_{\omega}^H G_{\omega}(\vec{r})|^2}{\|G_{\omega}(\vec{r})\|^2},
\end{equation}
(where $Y_{\omega}^H$ denotes the Hermitian transpose) which we will refer to as the {\em normalized ambiguity function}, and will refer to its maximization as {\em normalized Matched Field Processing} (nMFP). We show an example of the normalized ambiguity function in Fig.~\ref{fig:amb}.a.

The term $Y_{\omega}^H G_{\omega}(\vec{r})$ can be computed at every location $\vec{r}$ in an efficient manner using time-reversal.  Precise values for $\|G_{\omega}(\vec{r})\|^2$ are typically not available when computing the backpropagation $Y_{\omega}^H G_{\omega}(\vec{r})$. However it is often the case (and we will assume this here) that these energies either do not vary much across our locations of interest, or vary predictably (e.g.\ cylindrical spreading of the field amplitude).  Dropping the denominator yields the so-called {\em unnormalized ambiguity function} (alternatively the unnormalized Bartlett formulation) \cite{tolstoy2000applications,baggeroer1993overview,baggeroer1988matched}, the objective function used for estimating the source location:
\begin{eqnarray}
	\hat{\vec{r}} = \arg\max_{\vec{r}} |h(\vec{r})|^2 
	\quad\text{where}\quad h(\vec{r}) = Y_{\omega}^H G_{\omega}(\vec {r}),
\label{eq:amb}
\end{eqnarray}

\subsection{Broadband MFP}

Now suppose that most of the energy of the source signal occupies some continuous bandwidth $[\omega_{\text{min}} ~~ \omega_{\text{max}}]$, known as the broadband formulation. Ideally, we would solve \eqref{eq:ls-single} over a continuum of $\omega$ values. However, for the sake of source localization, it is computationally advantageous to sample this bandwidth at $K$ frequencies $\omega_1,\omega_2,\ldots,\omega_K$, yielding $K$ measurement vectors $Y_{\omega_k}$ where $k \in \{1,2,...K\}$. In this way, we can achieve a computational complexity at most $K$ times the single frequency case, without sacrificing much precision.

We now search for the location $\vec{r}$ that jointly matches the joint behavior of the measurements $Y_\omega$ over multiple frequencies $\omega_1,\omega_2,\ldots,\omega_K$.  The least-squares problem from \eqref{eq:ls-single} becomes
\begin{equation}
	\label{eq:ls-incoherent}
	\arg\min_{\vec{r}}\min_{\beta_{\omega_1},\ldots,\beta_{\omega_K}}
	\sum_{k=1}^K \|Y_{\omega_k} - \beta_{\omega_k}G_{\omega_k}(\vec{r})\|^2.
\end{equation}
The inner optimization problem is separable over the $\beta_{\omega_k}$, and so the above is equivalent to
\begin{equation}
	\arg\min_{\vec{r}}\sum_{k=1}^K\min_{\beta_{\omega_k}}\|Y_{\omega_k} - \beta_{\omega_k}G_{\omega_k}(\vec{r})\|^2
	~=~
	\arg\max_{\vec{r}}\sum_{k=1}^K \frac{|Y_{\omega_k}^HG_{\omega_k}(\vec{r})|^2}{\|G_{\omega_k}(\vec{r})\|^2}.
\label{eq:nIncoherent}
\end{equation}
As before, if the energies $\|G_{\omega_k}(\vec{r})\|^2$ are homogenous across space and frequency, then a reasonable unnormalized approximation to the above is 
\begin{equation}
	\arg\max_{\vec{r}} \sum_{k=1}^K |  Y_{\omega_k}^{H}  G_{\omega_k}(\vec{r}) |^2
	~=~ 
	\sum_{k=1}^K |h_{\omega_k}(\vec{r})|^2.
\label{eq:uIncoherent}
\end{equation}

\begin{figure*}
	\centering
	\begin{tabular}{cc}
		\raisebox{.5in}{\rotatebox{90}{Depth (m)}}
		\includegraphics[width=60.0mm]{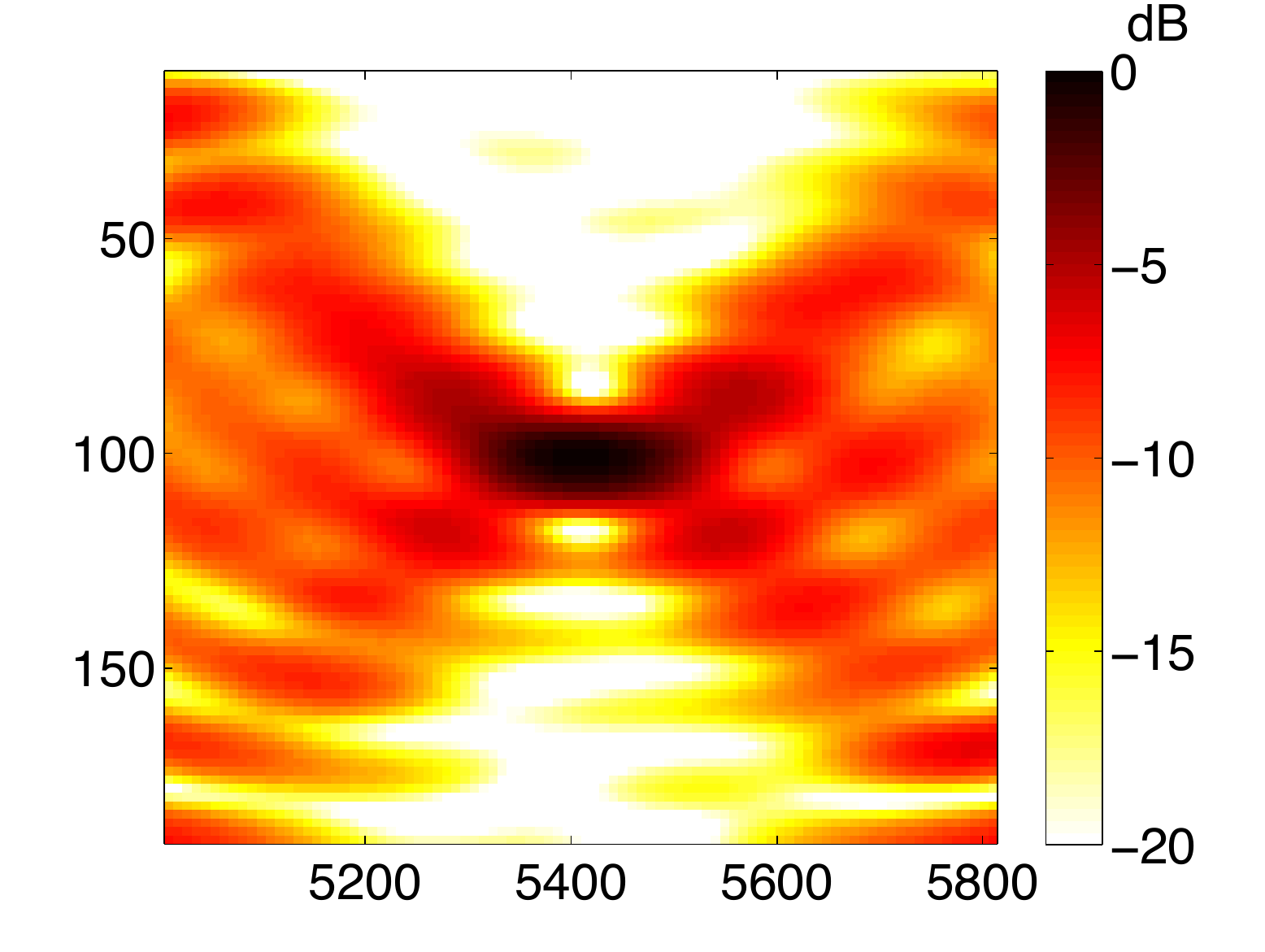} &
		\hspace{.1in} 
		\includegraphics[width=60.0mm]{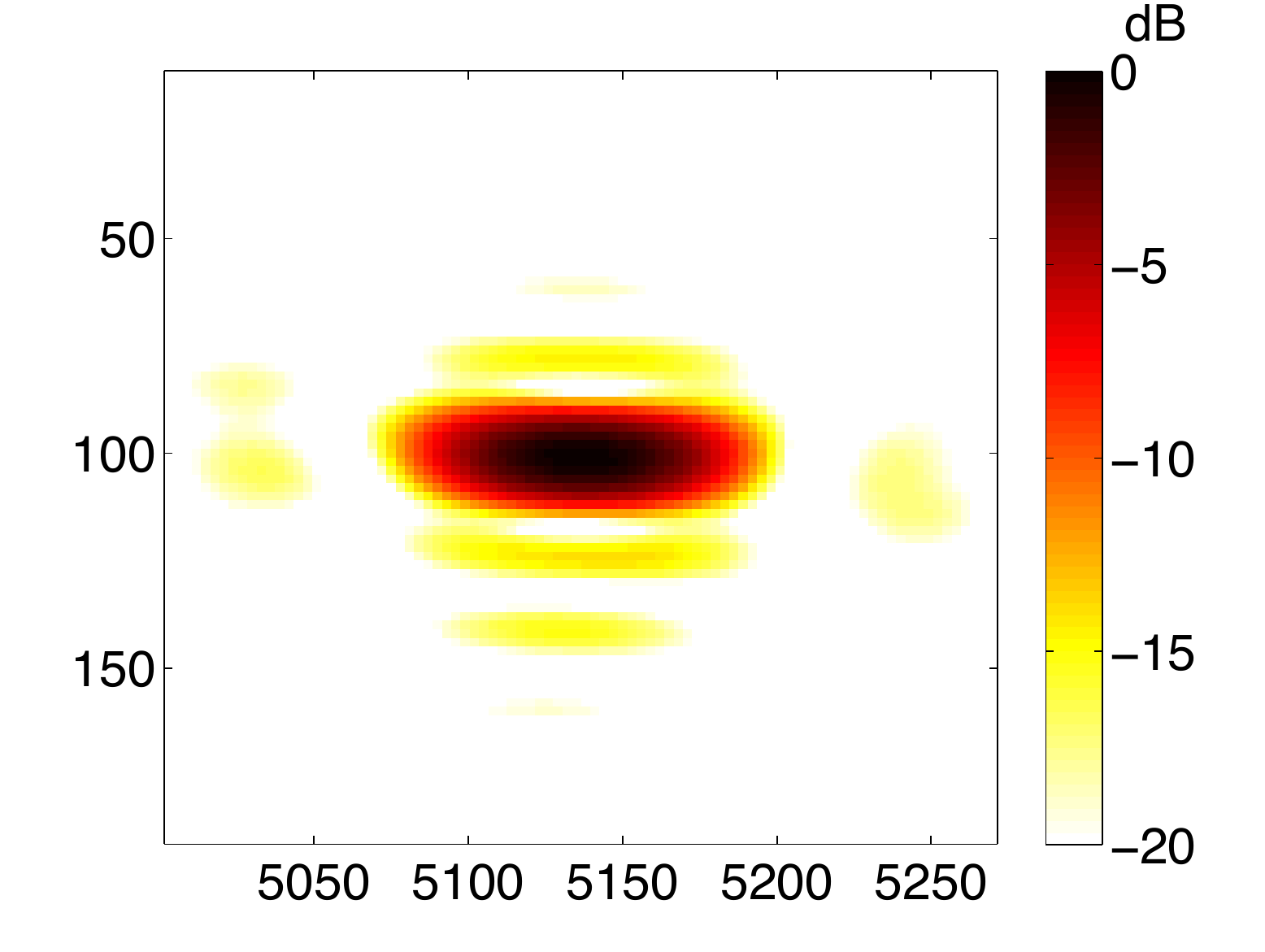} \\
		(a) & (b)
	\end{tabular}
	\centering
	\begin{tabular}{cc}
		\raisebox{.5in}{\rotatebox{90}{Depth (m)}}
		\includegraphics[width=60.0mm]{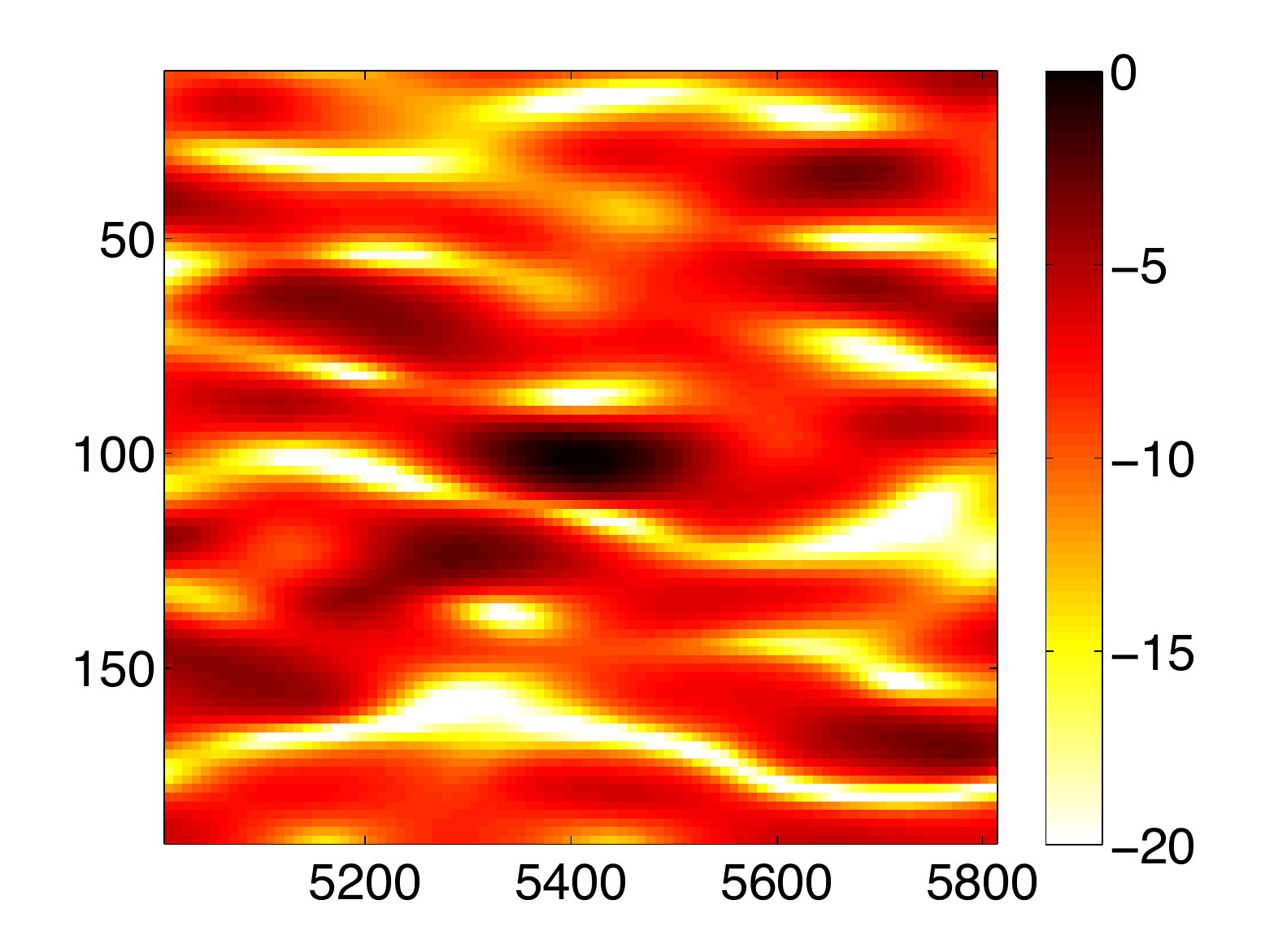} &
		\hspace{.1in} 
		\includegraphics[width=60.0mm]{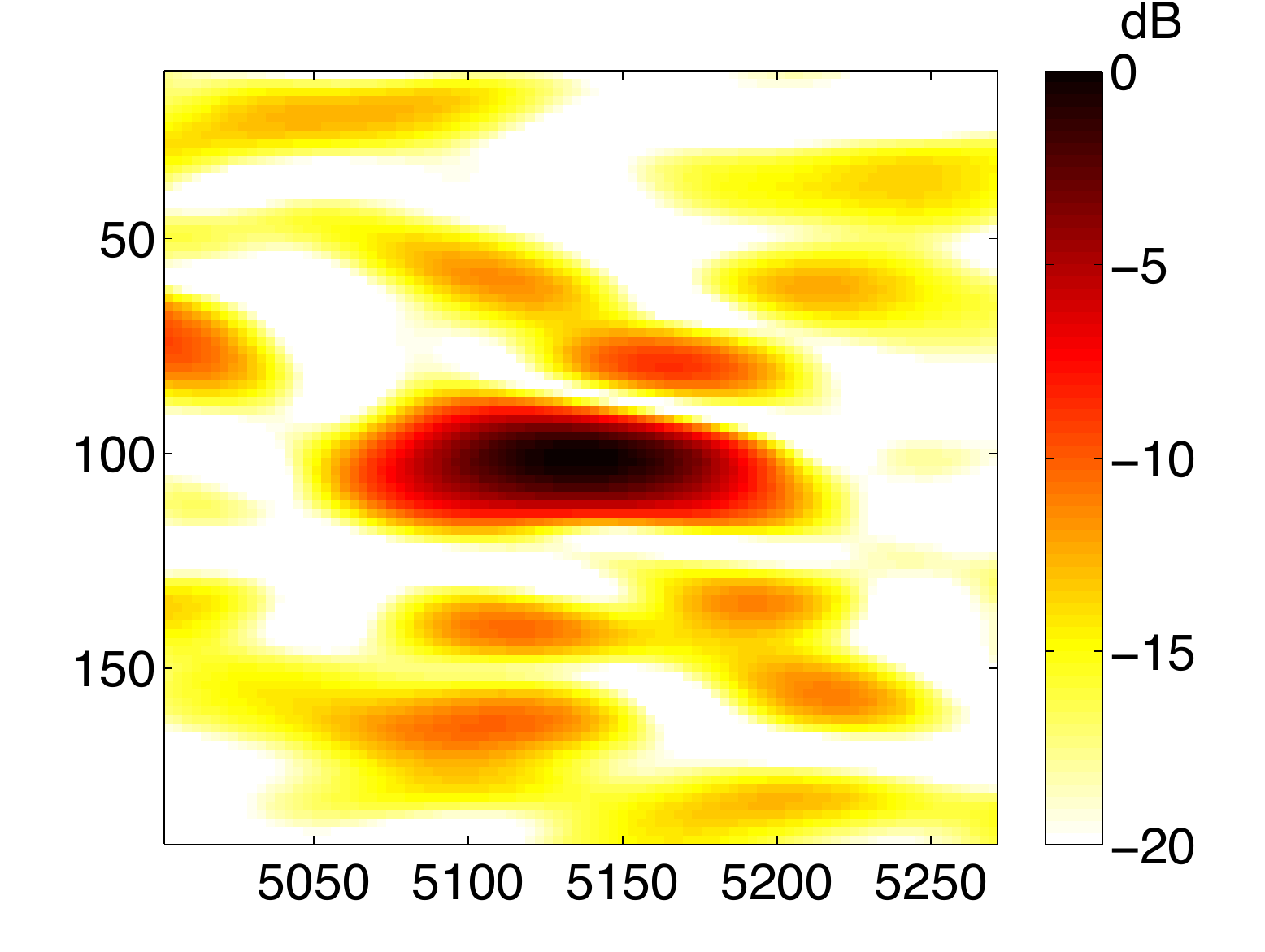} \\
		(c) & (d)
	\end{tabular}
	\centering
	\begin{tabular}{cc}
		\raisebox{.5in}{\rotatebox{90}{Depth (m)}}
		\includegraphics[width=60.0mm]{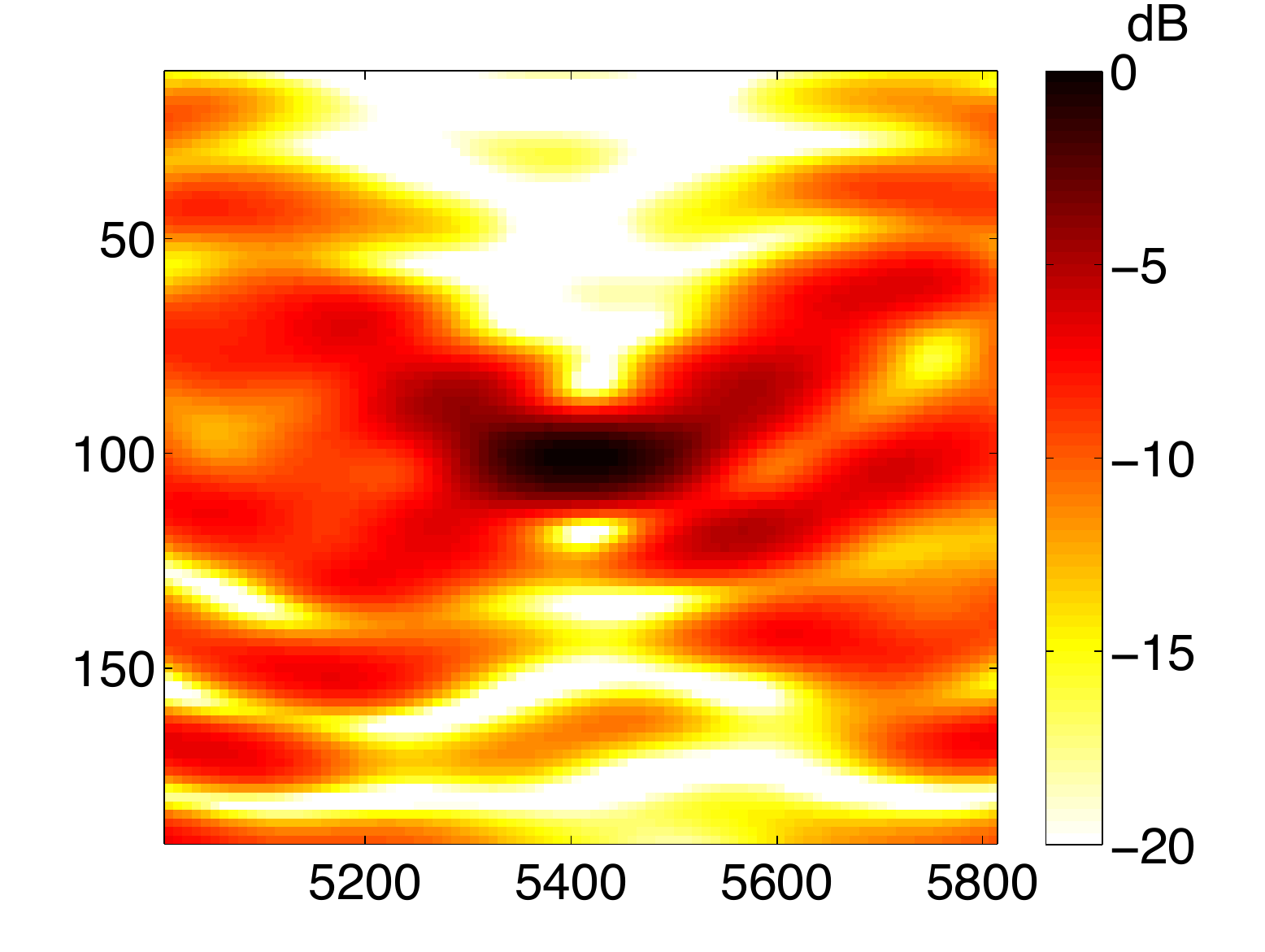} &
		\hspace{.1in} 
		\includegraphics[width=60.0mm]{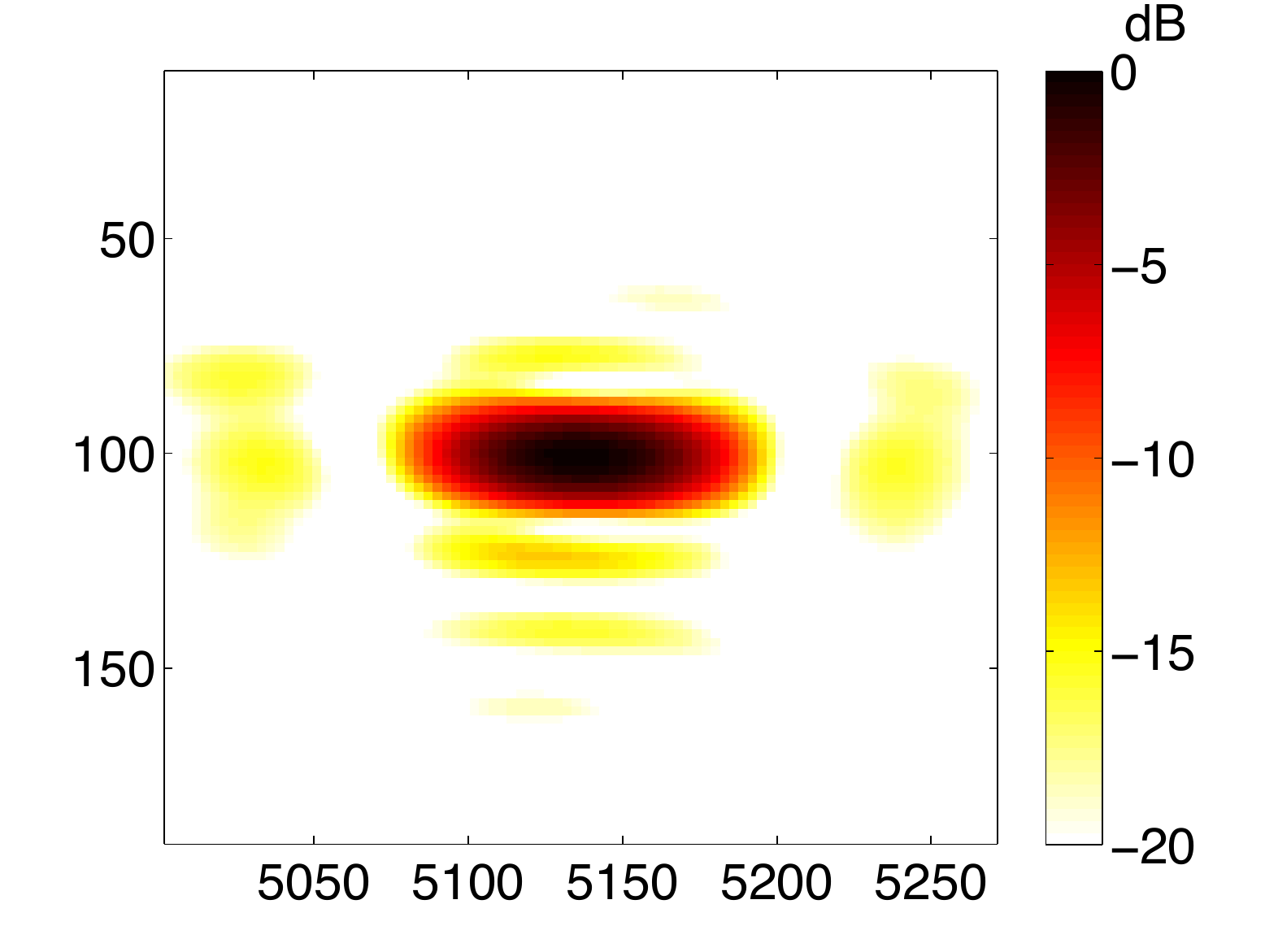} \\
		Range (m) & \hspace{.1in} Range (m) \\[1mm]
		(e) & (f)
	\end{tabular}
\caption{\small\sl Single-frequency (left column) and broadband-coherent (right column) ambiguity functions. These ambiguity functions shown on the dB scale ($20 \log_{10}(\cdot)$) for: (a, b) the standard MFP as described in Eqs.~\eqref{eq:amb-norm} and \eqref{eq:nCoherent}, and (c, d) cMFP as described in Eqs.~\eqref{eq:coded-mfp} and \eqref{eq:COHERENT_BROADBAND_CS_amb} for the single-frequency case with $M=10$ and broadband coherent case with $M=2$ measurements per frequency, and (e, f) cMFP for the single-frequency case with $M=30$ and broadband coherent case with $M=20$ measurements per frequency.
}
\label{fig:amb}
\end{figure*}

The formulation in \eqref{eq:uIncoherent} assumes that the source amplitudes $\beta_{\omega_k}$ are unknown.  If we have knowledge of the source signal's complex amplitudes, that is we know them up to a common amplitude and phase, then \eqref{eq:ls-incoherent} can be refined to
\begin{equation}
	\label{eq:ls-coherent}
	\arg\min_{\vec{r}}\min_{\beta\in\C}\left\|
	\begin{bmatrix}Y_{\omega_1}\\ Y_{\omega_2} \\ \vdots \\ Y_{\omega_K}\end{bmatrix} - 
	\beta\begin{bmatrix}\alpha_{\omega_1}G_{\omega_1}(\vec{r}) \\ \alpha_{\omega_2}G_{\omega_2}(\vec{r}) \\ \vdots \\ \alpha_{\omega_K}G_{\omega_K}(\vec{r})  \end{bmatrix}\right\|^2
\end{equation}
where the source amplitudes $\alpha_{\omega_k}$ are fixed and known.  Applying again the results from Appendix~\ref{app:closest-point}, the inner optimization program can be solved in closed form, and so \eqref{eq:ls-coherent} is equivalent to
\begin{equation}
	\arg\max_{\vec{r}}
	\frac{\left|\sum_{k=1}^K\alpha_{\omega_k}Y_{\omega_k}^HG_{\omega_k}(\vec{r})\right|^2}
	{\sum_{k=1}^K|\alpha_{\omega_k}|^2\|G_{\omega_k}(\vec{r})\|^2},
	\label{eq:nCoherent}
\end{equation}
as shown in Fig.~\ref{fig:amb}.b, which we can approximate (by removing the denominator) as its unnormalized counterpart
\begin{equation}
	\arg\max_{\vec{r}} \left|\sum_{k=1}^K \alpha_{\omega_k}h_{\omega_k}(\vec{r})\right|^2.
	\label{eq:uCoherent}
\end{equation}
Hereafter, we will refer to \eqref{eq:nIncoherent} and \eqref{eq:uIncoherent} as the {\em incoherent} MFP formulation, and \eqref{eq:nCoherent} and \eqref{eq:uCoherent} as the {\em coherent} MFP formulation.

\section{Compressive MFP}
\label{sec:cMFP}

In this section, we describe Compressive Matched Field Processing (cMFP). This is an efficient method for acquiring a {\em compressed} version of the Green's function operator $G_{\omega}(\vec{r})$ that exhibits a behavior in some regards similar to the dimension-reduced counterpart achieved via Principal Component Analysis, but may be obtained with only incomplete knowledge of the Green's function $G_{\omega}(\vect{r})$. Our approach works by precomputing the backpropagation of a small number of hypothetical received signals to construct a dimension-reduced proxy for the Green's function. Then, given the actual observed data $Y_{\omega}$ we localize the source by finding the closest match between the received signal and the Green's function in the compressed domain.  With the compressed version of $G_{\omega}(\vec{r})$ in hand, locating the source only requires computing a series of short inner products. In addition, the compressed version of $G_{\omega}(\vec{r})$ is independent of the received signal, and so can be pre-computed and re-used for later observations.  As we will demonstrate in Section~\ref{sec:results}, this cMFP strategy is effective even when the number of pre-computed \cms is far fewer than what would be required for a full acquisition of $G_{\omega}(\vect{r})$ over the whole search grid area.

\subsection{Single-Frequency cMFP}

We start by discussing the single-frequency case in detail.  First, we compute the {\em compressed} Green's function $\bPhi G_{\omega}(\vect{r})$, where $\bPhi$ is a $M\times N$ {\em encoding matrix}. Note that matrices are written in boldface letters in the remainder of this article. We construct  
$\bPhi G_{\omega}(\vect{r})$ by backpropagating (i.e.\ applying $G_{\omega}^H$ to) a series of {\em test vectors} $\Phi_1,\ldots,\Phi_M\in\C^N$ --- we will discuss how the $\Phi_m$ are chosen in the next section.

The result of one of these computations $\Phi_{m}^H G_{\omega}(\vec{r})$ is a complex-valued acoustic field over $\vec{r}$ and requires as much effort to compute as the ambiguity function $h(\vec{r})$.  We stack up the results of these precomputations as rows in the ensemble
\begin{equation}
	\begin{bmatrix} \Phi_{1}\herm G_{\omega}(\vec{r}) \\ 
		\Phi_{2}\herm G_{\omega}(\vec{r}) \\ \vdots \\ 
		\Phi_{M}\herm G_{\omega}(\vec{r}) 
	\end{bmatrix} 
	~=~ 
	\begin{bmatrix} \Phi_1 & \Phi_2 & ... & \Phi_M \end{bmatrix}^H G_{\omega}(\vec{r})
	~=~
	\bPhi G_{\omega}(\vec{r}).
\end{equation}
This ensemble gives us access to an indirect, dimension-reduced version of $G_{\omega}(\vec{r})$.

\begin{figure*}
\begin{tabular}{cc}
	\raisebox{27mm}{\rotatebox{90}{\Large dB}}
	\includegraphics[width=75.0mm]{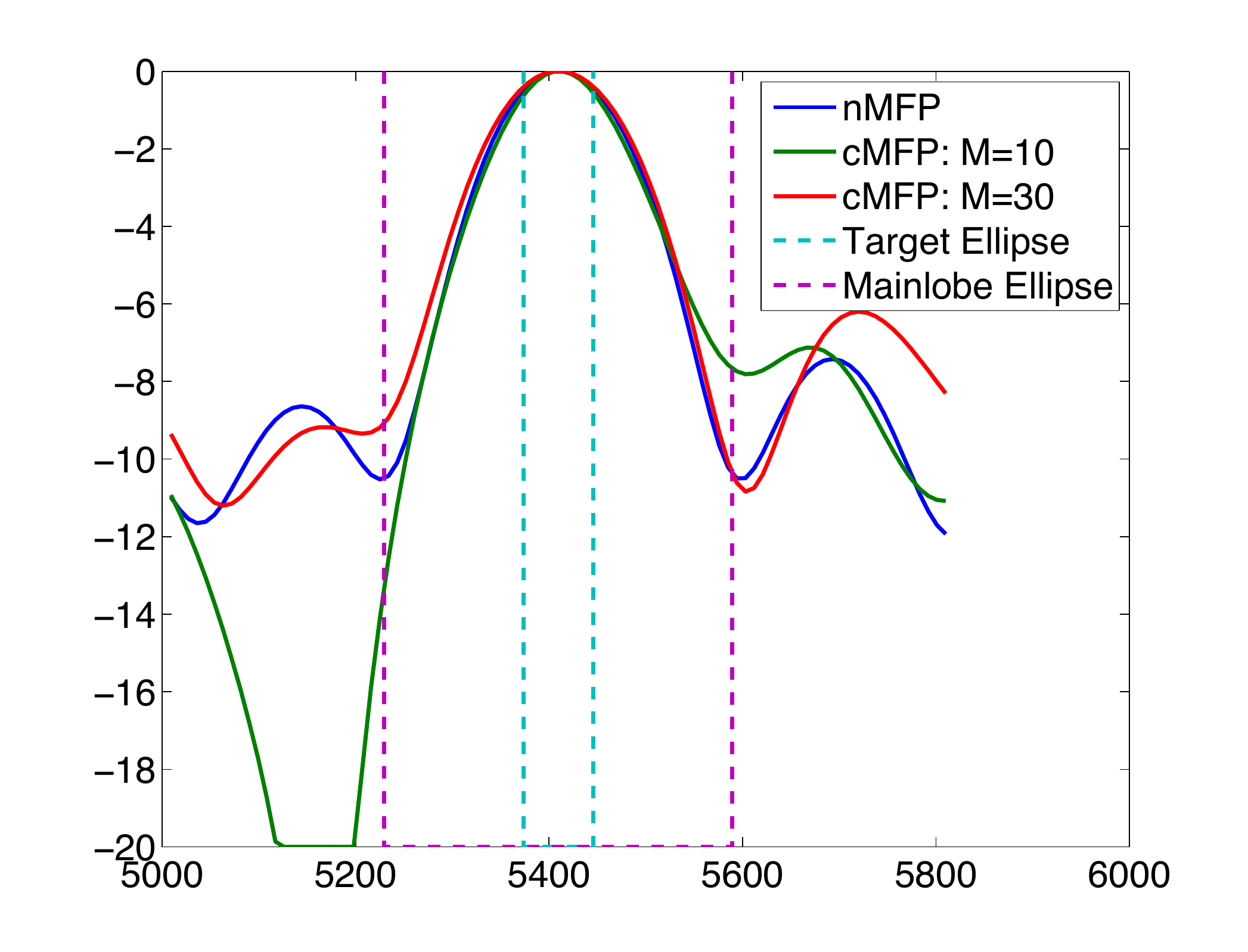} &
	\raisebox{27mm}{\rotatebox{90}{\Large dB}}
	\includegraphics[width=75.0mm]{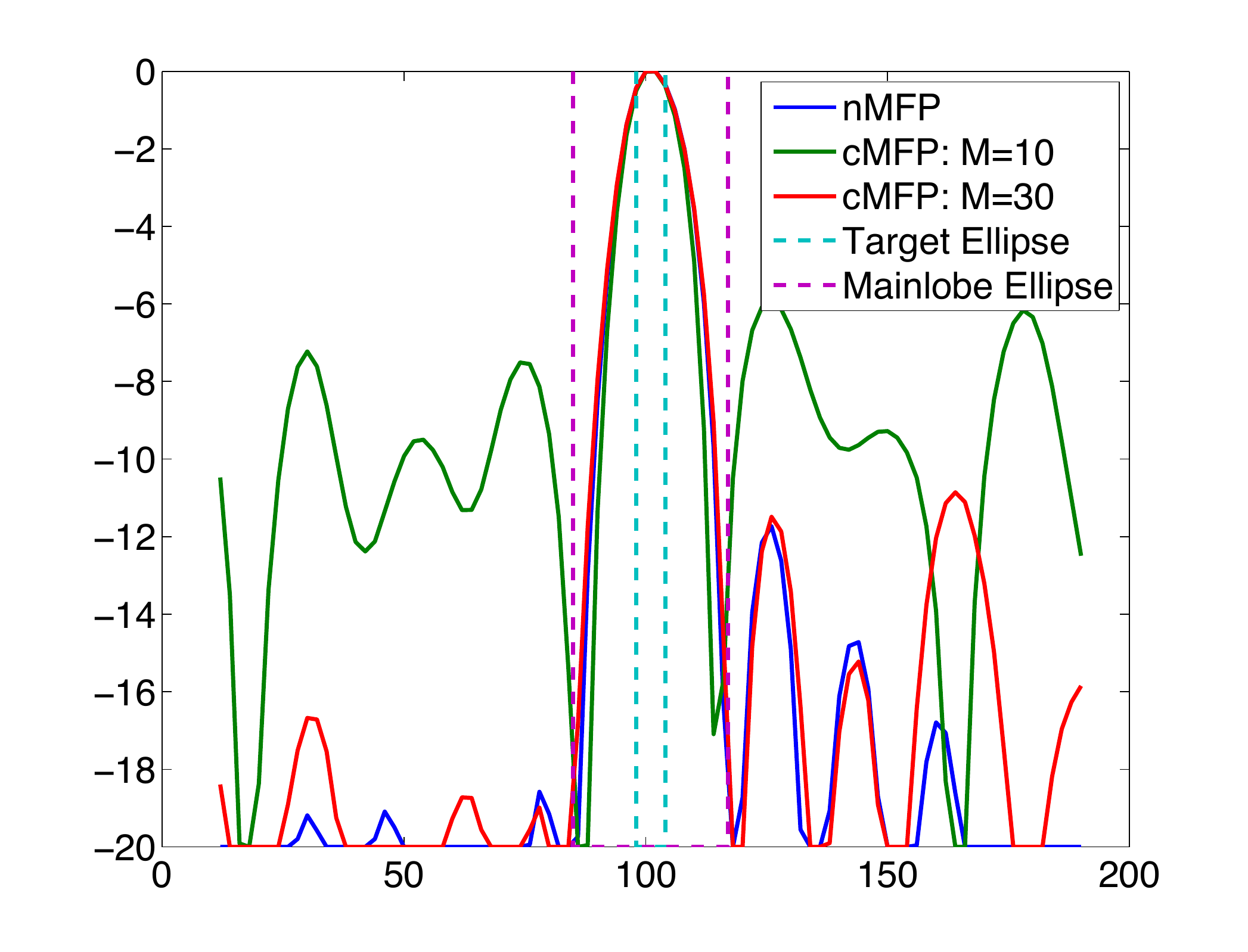}
	\\
	\hspace{.25in} range (m) &
	\hspace{.25in} depth (m) \\[.1in]
	(a) & (b)
\end{tabular}
\begin{tabular}{cc}
	\raisebox{27mm}{\rotatebox{90}{\Large dB}}
	\includegraphics[width=75.0mm]{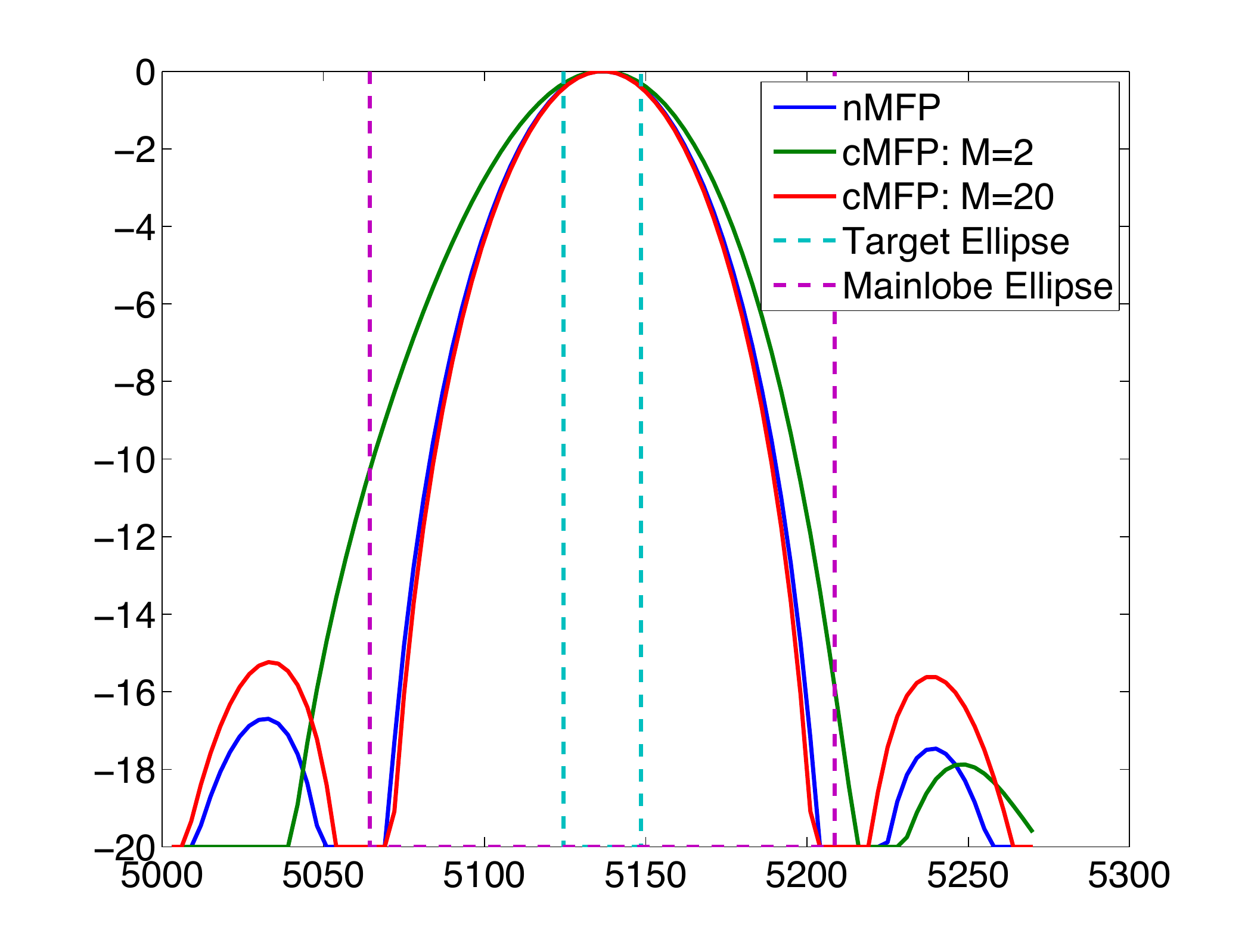} &
	\raisebox{27mm}{\rotatebox{90}{\Large dB}}
	\includegraphics[width=75.0mm]{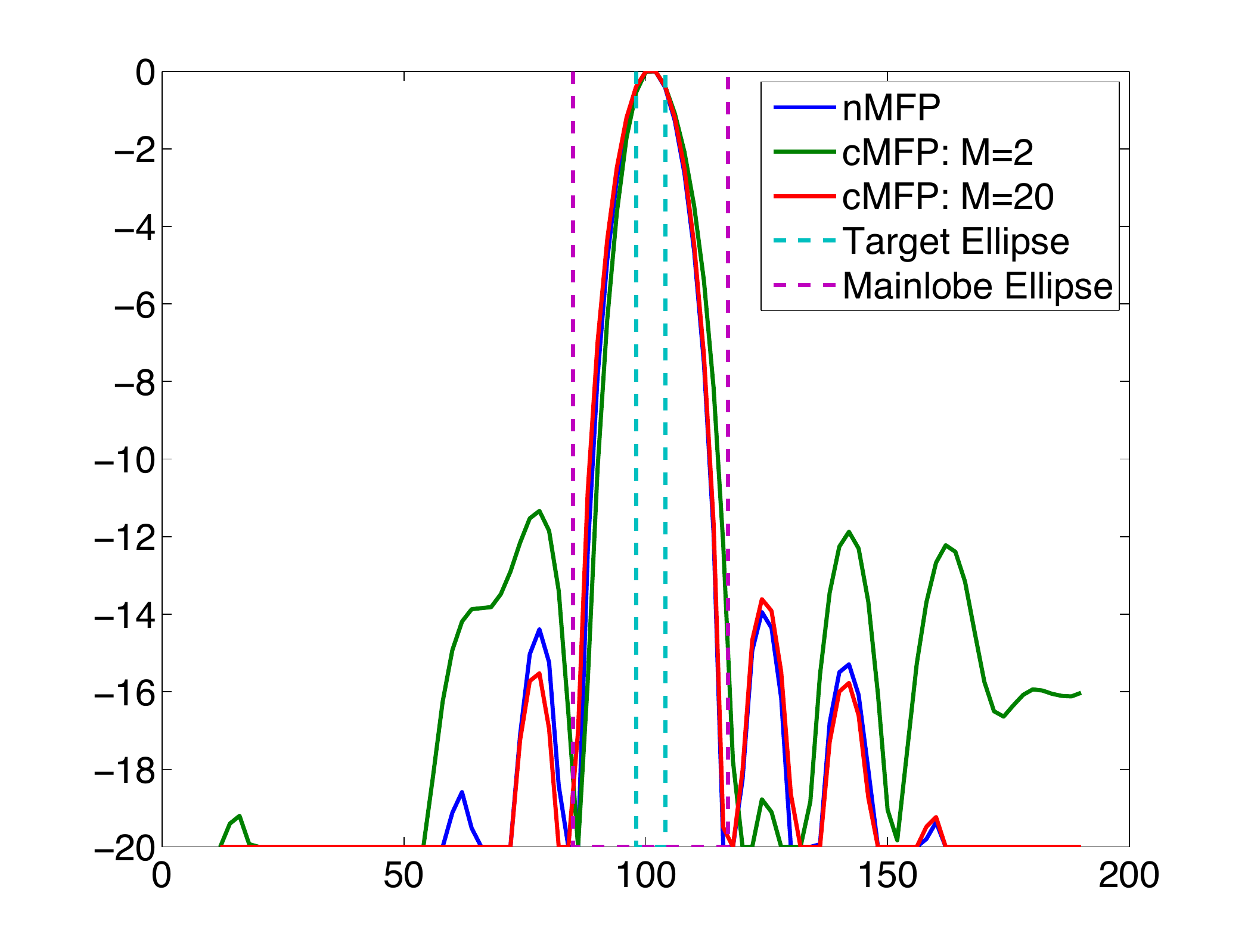}
	\\
	\hspace{.25in} range (m) &			
	\hspace{.25in} depth (m) \\[.1in]
	(c) & (d)
\end{tabular}
\caption{\small\sl  
Cross-sections of the ambiguity functions displayed on Fig. \ref{fig:amb}: (a, b) single-frequency case described by Eqs.~\eqref{eq:amb-norm} and Eqs.~\eqref{eq:coded-mfp}; (c, d) broadband coherent case Eqs.~\eqref{eq:nCoherent} and \eqref{eq:COHERENT_BROADBAND_CS_amb}; range (left column) and depth (right column). Here we show the normalized standard MFP (nMFP) and the cMFP (cMFP) for various values of $M$. The dashed lines show the boundaries for the main lobe and region of uncertainty that we are able to localize within under the presence of modest noise.}
\label{fig:cross}
\end{figure*}

Given observations $Y_{\omega}$, we search for the $\vec{r}$ that best explains these \cms in the compressed space.  The least-squares program \eqref{eq:ls-single} becomes
\begin{equation}
	\label{eq:ls-coded-single}
	\arg\min_{\vec{r}}\min_{\beta}~
	\|\bPhi Y_{\omega} - \beta\bPhi G_{\omega}(\vec{r})\|^2, 
\end{equation}
which, again using the results from Appendix~\ref{app:closest-point}, reduces to
\begin{equation}
	\label{eq:coded-mfp}
	\text{(narrowband cMFP)}\qquad
	\arg\max_{\vec{r}} \frac{|Y_{\omega}^H\bPhi^H\bPhi G_{\omega}(\vec{r})|^2}
	{\|\bPhi G_{\omega}(\vec{r})\|^2}
	\quad\text{where}\quad 
	\tilde{h}(\vec{r}) = Y_{\omega}^H\bPhi^H\bPhi G_{\omega}(\vec{r}).
\end{equation}
The function $\tilde{h}(\vec{r})$ is shown in Fig.~\ref{fig:amb}.c and \ref{fig:amb}.e, and can be interpreted as a compressed version of the ambiguity function $h(\vec{r})$ in \eqref{eq:amb} shown in Fig.~\ref{fig:amb}.a. The cross sections in range and depth of these ambiguity functions are shown in Fig.~\ref{fig:cross}.a and \ref{fig:cross}.b.

Note that unlike the standard MFP, in this case the precomputations give us direct access to the denominator $\|\bPhi G_{\omega}(\vec{r})\|^2$ (we simply take the norms of the columns of $\bPhi G_{\omega}(\vec{r})$), and so we leave it in the optimization program. As shown in the results section, this normalization term plays an important role in improving the source location estimation when the magnitude of the Green's function varies significantly across the search grid area.

Notice that an evaluation of \eqref{eq:coded-mfp} at a point $\vec{r}$ essentially only requires an inner product between the encoded observations $\bPhi Y_{\omega}$ and the $M$-vector $\bPhi G_{\omega}(\vec{r})$ formed from the backpropagated fields at point $\vec{r}$ from all $M$ test vectors.

\subsection{Random Projections}

The question remains as to how to choose the encoding matrix $\bPhi$ so that solution to the cMFP \eqref{eq:coded-mfp} is the same (or close to) the solution to the standard MFP \eqref{eq:amb}.  The corresponding least-squares problems are
\begin{eqnarray}
	\label{eq:MFP-ls-Yb}
	\text{standard MFP}: &\quad
	\arg\min_{\vec{r},\beta}
	\|Y_{\omega} - \beta G_{\omega}(\vec{r})\|^2 \\
	\label{eq:cMFP-ls-Yb}
	\text{cMFP}: &\quad
	\arg\min_{\vec{r},\beta}
	\|\bPhi\left(Y_{\omega} - \beta G_{\omega}(\vec{r})\right)\|^2.
\end{eqnarray}  
These two programs will have similar solutions if their functionals are close to one another for all values of $\beta$ and $\vec{r}$.  If $Y_{\omega}=\alpha G_{\omega}(\vec{r_0})$, then the performance of the cMFP will match that of the standard MFP when $\bPhi$ preserves the energy of the differences between the observations $Y_{\omega}$ and all scalar multiples of the Green's function at different points:
\begin{equation}
	\label{eq:embedding}
	\|\bPhi (F_1-F_2)\|^2 \approx \|F_1-F_2\|^2 \quad\text{for all}\quad
	F_1,F_2\in\mathcal{F}:= 
	\{F: F = \alpha G_\omega(\vec{r}),~\alpha\in\C;~\vec{r}\in\mathcal{R}\}.
\end{equation}
Essentially, we want $\bPhi$ to {\em stably embed} (i.e.\ preserve the distances between members of) the set $\mathcal{F}$ into $\C^M$.

We propose taking $\bPhi$ to be a random linear mapping.  This choice is inspired both by classical results in theoretical computer science and from the recently developed theory of compressive sensing.  In the mid-1980s, Johnson and Lindenstrauss \cite{johnson84ex} demonstrated that the distances within a finite set of $n$ points are essentially preserved through a random projection into a space of dimension $\sim\log n$ (see also \cite{dasgupta03el,achlioptas03da}).  Recently it has been shown that this same type of projection also embeds sparse signals into a low-dimensional subspace \cite{baraniuk08si}, a result which plays a key role in compressive sampling \cite{candes06ne,candes2006stable}, and are effective at reducing the dimensionality of certain types of manifolds \cite{baraniuk2009random}.  

We will discuss the particular the case where $\bPhi$ is a {\em random orthoprojection}, although the results will be almost identical for many different choices of random $\bPhi$ (e.g. with entries that are independent and identically distributed Gaussian or $\pm 1$ random variables). To generate $\bPhi$, we simply draw an $M\times N$ matrix of independent Gaussian random variables with unit variance, orthonormalize the rows using the Gram-Schmidt (or QR) algorithm, and then multiply by $\sqrt{N/M}$.  For an arbitrary fixed vector $F$, the random orthoprojection $\bPhi$ obeys two properties\cite{dasgupta03el}:
\begin{align}
	\E{\|\bPhi F\|^2} &= \|F\|^2 \\
	\label{eq:concentration}
	\P{\left|\, \|\bPhi F\|^2 - \|F\|^2\,\right| > \epsilon} &\leq 
	2e^{-\frac{M}{2\|F\|^2}\left(\epsilon^2/2 - \epsilon^3/3\right)}.
\end{align}
This allows us to interpret the compressed energy functional $\|\bPhi(Y_{\omega}-\beta G_{\omega}(\vec{r}))\|^2$ in \eqref{eq:cMFP-ls-Yb} as a random process, indexed by $\beta$ and $\vec{r}$, whose mean is the standard energy functional $\|Y_{\omega}-\beta G_{\omega}(\vec{r})\|^2$ in \eqref{eq:MFP-ls-Yb}.  At a fixed point $\beta,\vec{r}$, this random process is concentrated around its mean roughly like a Gaussian random variable with standard deviation $\sqrt{2/M}\|Y_{\omega}-\beta G_{\omega}(\vec{r})\|$.  The larger we make $M$ (the more random vectors we precompute backpropagations for), the tighter the concentration.  By construction, when $M=N$, $\bPhi^H\bPhi = I$ and we have acquired a ``lossless'' version of the Green's function $G_{\omega}(\vec{r})$, meaning that the functionals are exactly equal to one another.  In general, however, we will be interested in cases where there is a significant compression factor $M\ll N$ and benefit from the associated computational savings.

\subsection{Broadband cMFP}

The cMFP formulation can be readily extended to combine observations at multiple frequencies in both the incoherent and coherent cases.  For frequencies $\omega_1,\omega_2,\ldots,\omega_K$, we generate a sequence of $M\times N$ random matrices $\bPhi_{\omega_1},\bPhi_{\omega_2},\ldots,\bPhi_{\omega_K}$ and backpropagate the rows of each (for a total of $MK$ time-reversals) to acquire $\bPhi_{\omega_1} G_{\omega_1}(\vec{r}),\ldots,\bPhi_{\omega_K} G_{\omega_K}(\vec{r})$.  Then given observations $Y_{\omega_1},\ldots,Y_{\omega_K}$, we compress them by calculating $\bPhi_{\omega_1}Y_{\omega_1},\ldots,\bPhi_{\omega_K}Y_{\omega_K}$, and then using the compressed versions of the $G_{\omega_k}$, we proceed as in \eqref{eq:ls-incoherent} for the incoherent case
\begin{align}
	\label{eq:INCOHERENT_BROADBAND_CS_amb}
	\text{(incoherent cMFP)}\qquad &
	\arg\min_{\vec{r}}\min_{\beta_{\omega_1},\ldots,\beta_{\omega_K}} 
	\sum_{k=1}^K\|\bPhi_{\omega_k} Y_{\omega_k} - \beta_{\omega_k}\bPhi_{\omega_k} G_{\omega_k}(\vec{r})\|^2 \nonumber \\
	&= \arg\min_{\vec{r}}\sum_{k=1}^K\min_{\beta_{\omega_k}}
	\|\bPhi_{\omega_k} Y_{\omega_k} - \beta_{\omega_k}\bPhi_{\omega_k} G_{\omega_k}(\vec{r})\|^2 \nonumber \\
	&= \arg\max_{\vec{r}}\sum_{k=1}^K 
	\frac{|Y_{\omega_k}^H\bPhi_{\omega_k}^H\bPhi_{\omega_k} G_{\omega_k}(\vec{r})|^2}
	{\|\bPhi_{\omega_k} G_{\omega_k}(\vec{r})\|^2},
	\nonumber \\
	&= \arg\max_{\vec{r}}\sum_{k=1}^K 
	\frac{|\tilde{h}_{\omega_k}(\vec{r})|^2}{\|\bPhi_{\omega_k} G_{\omega_k}(\vec{r})\|^2}
\end{align}
and as in \eqref{eq:ls-coherent} for the coherent case:
\begin{align}
	\label{eq:COHERENT_BROADBAND_CS_amb}
	\text{(coherent cMFP)}\qquad &
	\arg\min_{\vec{r}}\min_{\beta}\left\|\begin{bmatrix}
	\bPhi_{\omega_1} Y_{\omega_1} \\ \bPhi_{\omega_2} Y_{\omega_2} \\ \vdots \\ 
	\bPhi_{\omega_K} Y_{\omega_K}\end{bmatrix}
	- \beta\begin{bmatrix}
	\alpha_{\omega_1}\bPhi_{\omega_1} G_{\omega_1}(\vec{r}) \\ \alpha_{\omega_2}\bPhi_{\omega_2} G_{\omega_2}(\vec{r}) \\
	\vdots \\ \alpha_{\omega_K}\bPhi_{\omega_K} G_{\omega_K}(\vec{r}) \end{bmatrix}
	\right\|^2 \nonumber \\[3mm]
	&= \arg\max_{\vec{r}}\frac{\left|\sum_{k=1}^K
	\alpha_{\omega_k} Y_{\omega_k}\bPhi_{\omega_k}^H\bPhi_{\omega_k} G_{\omega_k}(\vec{r})\right|^2}
	{\sum_{k=1}^K|\alpha_{\omega_k}|^2\ \|\bPhi_{\omega_k} G_{\omega_k}(\vec{r})\|^2} \nonumber \\[3mm]
	&= \arg\max_{\vec{r}} 
	\frac{\left|\sum_{k=1}^K\alpha_{\omega_k}\tilde{h}_{\omega_k}(\vec{r})\right|^2}
	{\sum_{k=1}^K |\alpha_{\omega_k}|^2\ \|\bPhi_{\omega_k} G_{\omega_k}(\vec{r})\|^2}.
\end{align}
The incoherent and coherent case are respectively illustrated in Fig.~\ref{fig:amb}.d and \ref{fig:amb}.f and in Fig.~\ref{fig:cross}.c and \ref{fig:cross}.d. Note that in this coherent case, the optimization is identical in its structure to the single-frequency case. In particular, by concatenating:
\begin{equation}
G(\vect{r}) = 
\begin{bmatrix}
	\alpha_{\omega_1} G_{\omega_1}(\vec{r}) \\ 
	\alpha_{\omega_2} G_{\omega_2}(\vec{r}) \\
	\vdots \\ 
	\alpha_{\omega_K} G_{\omega_K}(\vec{r}) 
\end{bmatrix}
\quad
Y = \begin{bmatrix} Y_{\omega_1} \\ Y_{\omega_2} \\ \vdots \\  Y_{\omega_K} \\ \end{bmatrix}
\quad
\bPhi = 
\begin{bmatrix}
\bPhi_{\omega_1} & & & \\
& \bPhi_{\omega_2} & & \\
& & \ddots & \\
& & & \bPhi_{\omega_K}
\end{bmatrix},
\end{equation}
we see that the coherent broadband formulation \eqref{eq:COHERENT_BROADBAND_CS_amb} shares the same formulation as the single frequency case \eqref{eq:coded-mfp}.


\section{Numerical simulations}
\label{sec:results}


In this section, we present numerical experiments demonstrating that underwater acoustic sources can be localized from highly compressed versions of the Green's functions.  Our cMFP results give locations estimates for single-frequency, incoherent broadband, and coherent broadband that are comparable with the traditional MFP.  After the initial pre-computation (which consists of backpropagating the random codes at each frequency), the cMFP is substantially faster than the traditional MFP, requiring only a short inner product to be calculated at each search location. 

The MATLAB code generating all the numerical results presented in this section is available online 
\footnote{Download the code at \href{http://users.ece.gatech.edu/~wmantzel3/cmfp/code.zip}{http://users.ece.gatech.edu/\textasciitilde wmantzel3/cmfp/code.zip}.}.

\vspace{.2in}
\subsection{Numerical set-up}
All numerical simulations were conducted using a $200$m deep Pekeris waveguide and the Green's functions were computed using a standard normal mode code \cite[pg 540--552]{jensen1994computational}. The two dimensional search grid area in depth and range spans respectively  $[10{\rm m}~~ 190{\rm m}]$, and  $[5000{\rm m}~~5810{\rm m}]$  for the single frequency and broadband incoherent simulations. The range span for the  broadband coherent simulations was reduced to $[5000{\rm m}~~5270{\rm m}]$ to keep constant the number of search locations  over which the ambiguity functions are computed since the effective resolution of the ambiguity function in the coherent case was  about $3$ times higher in range (see Fig.~\ref{fig:amb} and Fig.~\ref{fig:cross}).  A uniformly spaced vertical line array with $N=37$ elements spaced between $10$ and $190$ meters was used to sample the acoustic field. The Green's functions  between each of the search locations and the receiver array (see Fig. \ref{fig:LaMer}) were calculated across $K=20$ different frequencies between $141$ Hz and $160$ Hz (the narrowband configuration uses $150$ Hz). Given the selected numerical set-up, the natural resolution in frequency of the computed Green's function is around $5$ Hz; that is, $G_{\omega_1}(\vec{r})$ and $G_{\omega_2}(\vec{r})$ are essentially uncorrelated when $|\omega_1-\omega_2|\geq 10\pi$.  The selected sample spacing of $1$ Hz falls well within this frequency resolution.

After selected a source location $\vec{r_0}$ inside the region of interest, observations at the $K$ frequencies were simulated using the forward model, and uncorrelated zero-mean Gaussian noise was added to the result:
\begin{equation}
	Y_{\omega_k} = \alpha_{\omega_k} G_{\omega_k}(\vec{r_0}) + Z_k,\quad
	Z_k\in\C^N,~~Z_k\sim\mathrm{Normal}(0,\sigma^2I),
\label{Noisy_Data}
\end{equation}
where each $Z_k$ has i.i.d. Gaussian real and imaginary parts with variance $\sigma^2/2$. In all of our experiments, we set $\alpha_{\omega_k} = 1$ for all $k$.  The signal-to-noise ratio (SNR) corresponding to noise variance $\sigma^2$ is 
\begin{equation}
	\text{SNR}~=~ 10\log_{10}
	\left(\frac{|\alpha_{\omega}|^2\|G_{\omega}(\vec{r_0})\|^2}{N\sigma^2}\right)
\end{equation}
in the single frequency case, and
\begin{equation}
	\text{SNR}~=~ 10\log_{10}
	\left(\frac{\sum_{k=1}^K|\alpha_{\omega_k}|^2\ \|G_{\omega_k}(\vec{r_0})\|^2}{KN\sigma^2}\right)
\end{equation}
in the broadband case.  Unless otherwise stated, we used an SNR of $16$ dB.

\vspace{.1in}
\noindent

Given a set of observations, we estimate the source location by solving \eqref{eq:coded-mfp} single frequency), \eqref{eq:INCOHERENT_BROADBAND_CS_amb} (broadband incoherent), or \eqref{eq:COHERENT_BROADBAND_CS_amb} (broadband coherent) and compare against the standard MFP formulations \eqref{eq:amb}, \eqref{eq:uIncoherent}, and \eqref{eq:uCoherent} 
 As stated, these optimizations problems are over a continuous variable $\vec{r}$ --- in practice, we compute these functionals on a finite grid of points and choose the maximum from amongst these points.  We used a $90\times 90$ grid for the simulations presented below, which corresponds to $2$m spacing in depth, a $9$m spacing in range in the single-frequency and broadband incoherent cases, and $3$m  spacing in range in the broadband coherent case.
We wish to emphasize that while our solution will of course lie on one of these grid points, the actual source location is chosen to be an arbitrary point.

\begin{figure*}
	\centering
	\begin{tabular}{c}
		\raisebox{30mm}{\rotatebox{90}{\bf Depth (m)}}
		\includegraphics[width=100.0mm]{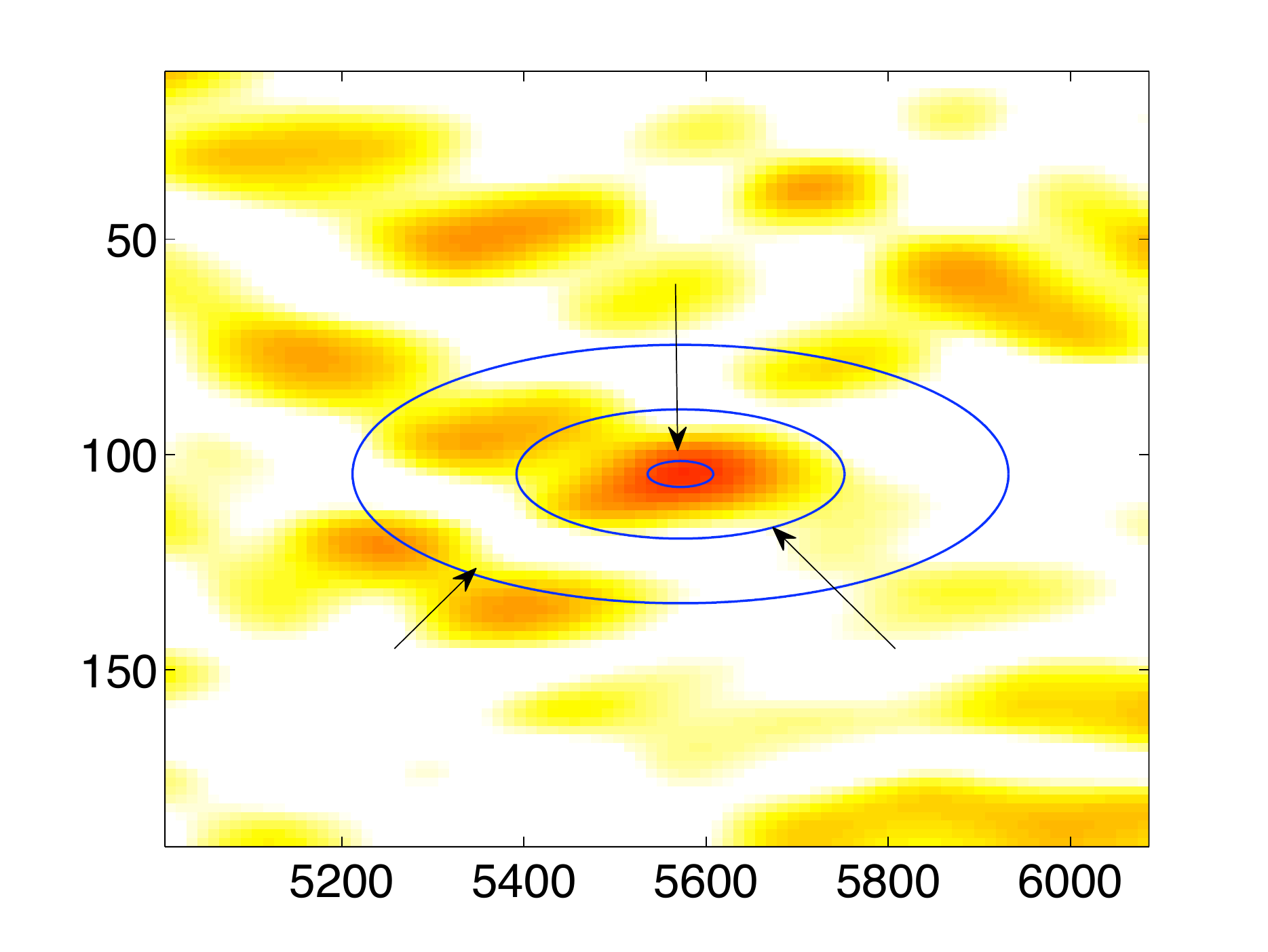}   \\[-60mm]
		 \hspace{-00mm} { $\|\vect{r}_0 - \vect{r}\|_e = 1$} \\[31mm]
		 \hspace{10mm} { $\|\vect{r}_0 - \vect{r}\|_e = 10$} 
		 \hspace{20mm} { $\|\vect{r}_0 - \vect{r}\|_e = 5$} \\[19mm]
		  { \bf Range (m)}
	\end{tabular}
	\caption{\small\sl 
Definition of the elliptical distance metric used for the performance study of cMFP. Because range errors tend to be greater than depth errors in long-range localization estimates, our results use an elongated distance metric that gives greater weight to the depth than the range, leading to unit balls that are ellipses instead of circles as shown here (see Eq. \eqref{eq:elong}). The color scheme for this ambiguity surface $10 \log_{10}(|h(\vect{r})|^2) = 10 \log_{10}(|Y_\omega^H G_\omega(\vec{r})|^2)$ has been lightened somewhat to allow for better visibility for the overlaid ellipses. 
}
\label{fig:elong}
\end{figure*}

The natural resolutions in depth and range of the ambiguity function $h_\omega(\vec{r})$ differ, as shown in Fig.~\ref{fig:elong} and Fig.~\ref{fig:cross} for a source located at $\vec{r_0} = (5540m,100m)$ in (range,depth)  for a single frequency $\omega = 300\pi$ rad/sec (150 Hz) for a source located at $\vec{r_0} = (5540,100)$ in (range,depth). In this case, the main lobe has a width of $\sim 360$ m in range and $\sim 32$ m in depth.  Again the grid spacing of $9m$/$2m$ in range/depth falls well within this resolution. The spatial resolution of the ambiguity surface in the selected multi-modal Pekeris waveguide is primarily a function of the source-receiver array configuration as well as the selected frequency band  \cite{tolstoy2000applications,baggeroer1993overview}
%
In light of these differing spatial resolutions, we use a weighted norm to report distance errors in most cases presented here in this section. The distance from a point $\vec{r_0} =  (r^\mathrm{range}_0,r^{\mathrm{depth}}_0)$ to the estimated source location $\hat{\vec{r}}$ is computed using the elliptical distance:
\begin{equation}
	\|\vec{r_0} - \hat{\vec{r}}\|_{\mathrm{e}} = 
	\sqrt{\left(\frac{r_0^\mathrm{range}-\hat{r}^\mathrm{range}}{e^\mathrm{range}}\right)^2 + 
	\left(\frac{r_0^\mathrm{depth}-\hat{r}^\mathrm{depth}}{e^\mathrm{depth}}\right)^2}.
	\label{eq:elong}
\end{equation}
We use $e^\mathrm{depth} = 3$m and $e^\mathrm{range}=36$m for the single-frequency and incoherent cases, and $e^\mathrm{range}=12$m in the coherent case.  The values of $e^\mathrm{depth}$ and $e^\mathrm{range}$ were chosen so that the contour $\{\vec{r}:~\|\vec{r_0}-\vec{r}\|_\mathrm{e}=1\}$ was approximately the same as the isosurface of the ambiguity function at $0.9$ of its maximum.
Equidistant points from $\vec{r_0} = (5540,100)$ for $\|\vec{r_0} - \vec{r}\|_\mathrm{e} =1,5$, and $10$ are shown in Fig.~\ref{fig:elong}.  For example, an error of $14.4$ meters in range and $0.9$ meters in depth translates to $0.5$ units of distance error in the elleptical $\|\cdot\|_\mathrm{e}$ norm.

\vspace{.2in}
\noindent
\subsection{Localization performance of cMFP.}  

\begin{figure*}
	\centering
	\begin{tabular}{cc}
		\raisebox{0.5in}{\rotatebox{90}{\Large $\Pn_M \left\{\text{error} > d\right\}$}}
		\includegraphics[width=70.0mm]{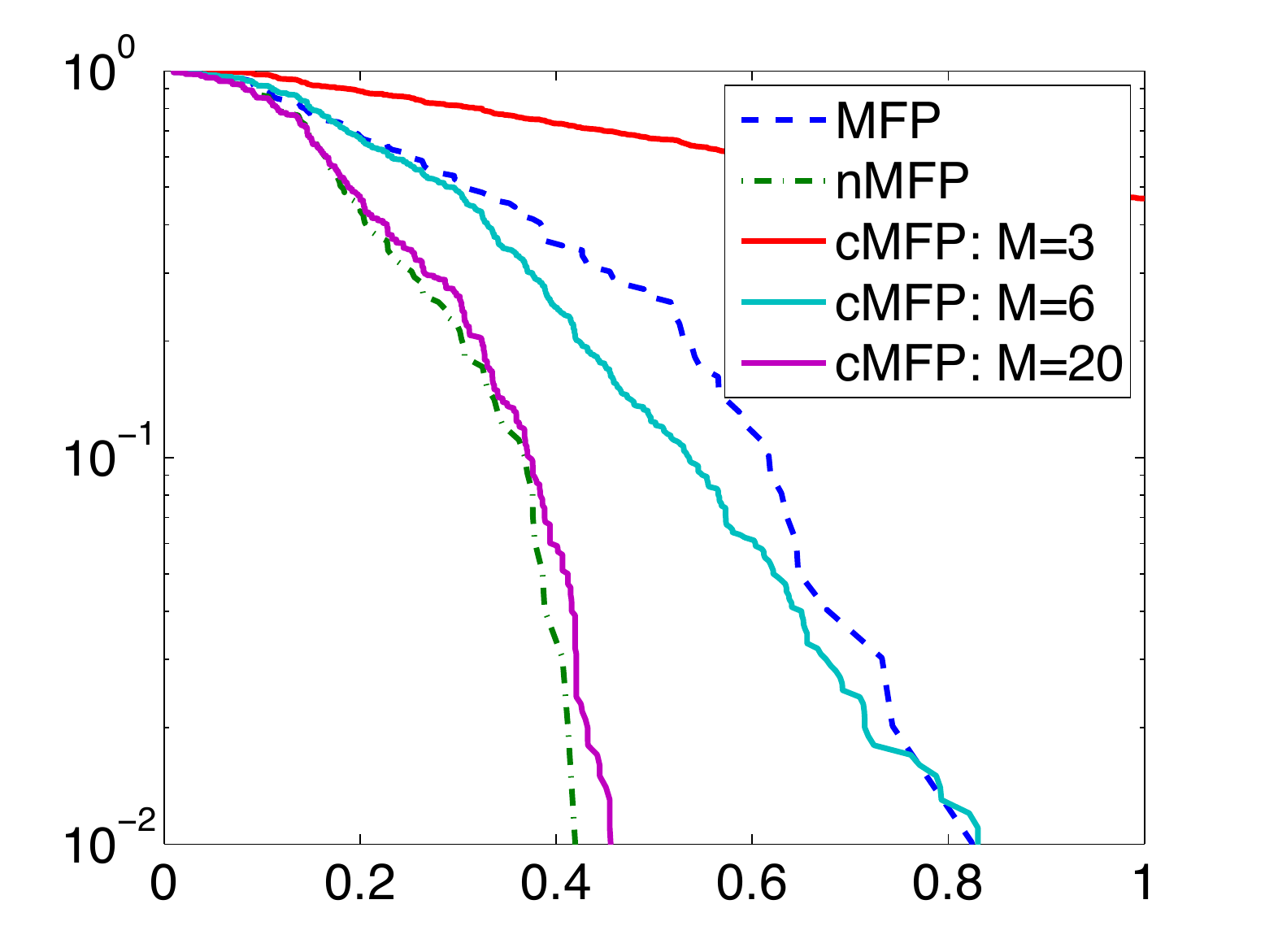} \\
		\hspace{.25in} Distance $d$ (in units of target ellipse radii) \\ (a) \\
		\raisebox{0.2in}{\rotatebox{90}{\Large $\Pn_M \left\{\text{error} > d=1\right\}$}}
		\includegraphics[width=70.0mm]{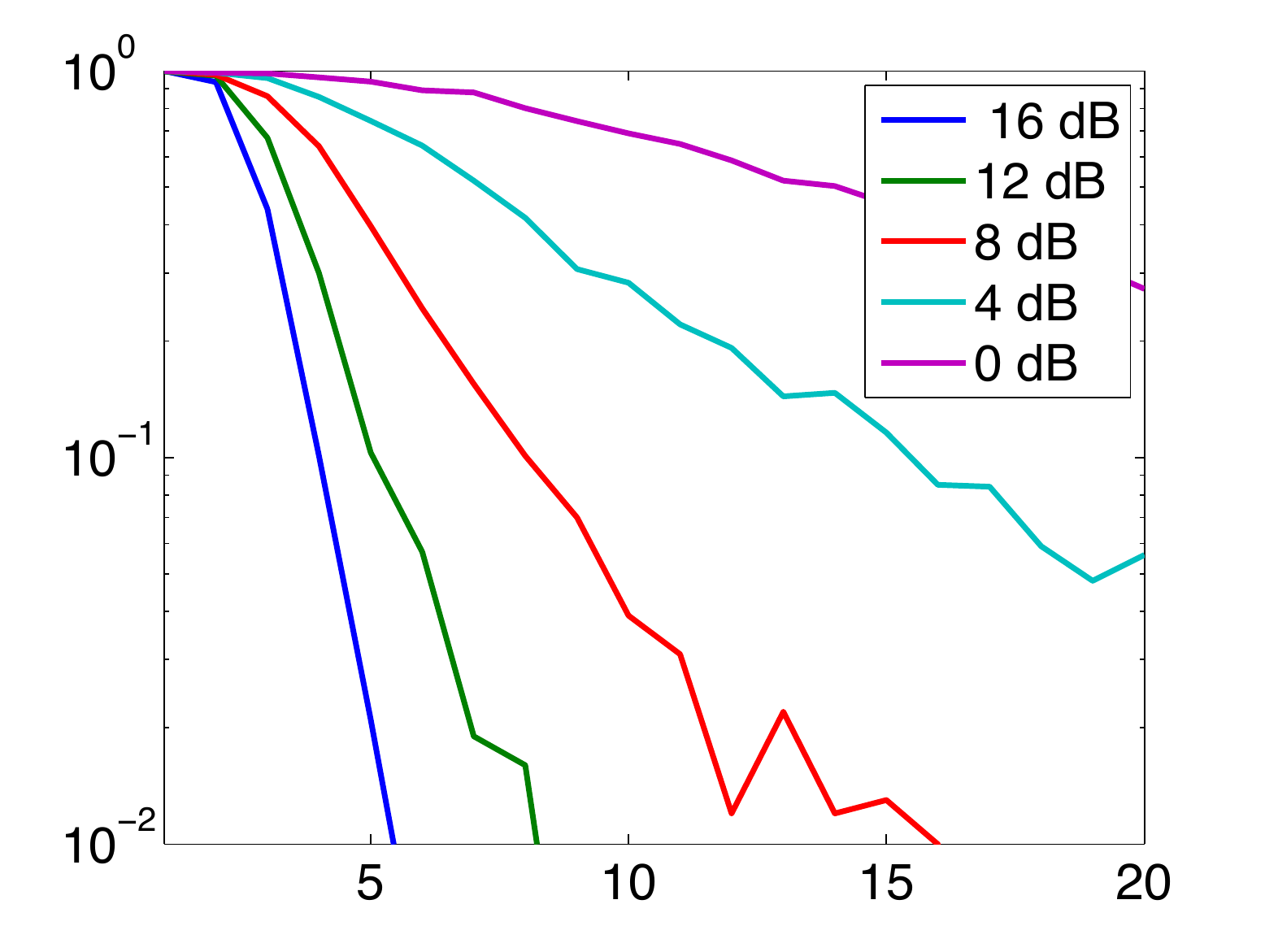} \\
		\hspace{.25in} Random Backpropagations: $M$ \\ (b) \\
		\raisebox{0.2in}{\rotatebox{90}{\Large $\Pn_{M=20} \left\{\text{error} > d\right\}$}}
		\includegraphics[width=70.0mm]{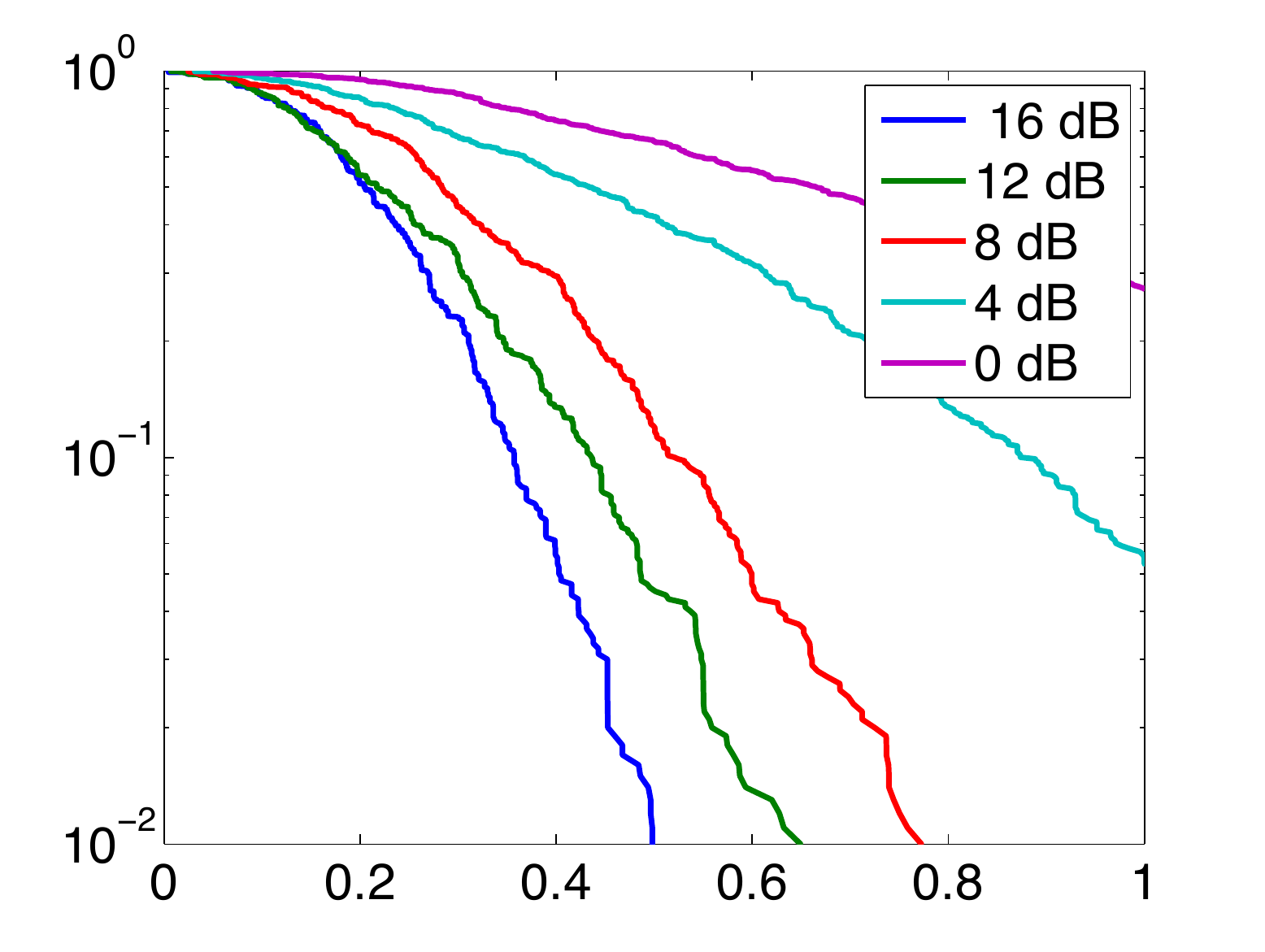} \\
		\hspace{.25in} Distance $d$ (in units of target ellipse radii) \\ (c)
	\end{tabular}
	\caption{\small\sl  
	(a) Tail probability of distance error $\|\hat{\vect{r}} - \vect{r_0}\|_e$ (see Eq. \eqref{eq:elong}) for the single-frequency cMFP formulation (see Eq. \eqref{eq:coded-mfp}) at $150$ Hz. $P_M(d)$ is the probability that the localization is worse than some distance $d$ using $M$ \cms. The dashed lines indicate the performance under normalized and unnormalized MFP (Eq. \eqref{eq:amb-norm} and Eq. \eqref{eq:amb}).
	The next two plots show results for $P_M(d)$ over various SNRs of the received signal with (b) fixing $d=1$ and (c) fixing $M=20$.
	}
\label{fig:single}
\end{figure*}

Fig.~\ref{fig:single}a compares the performance of cMFP (see Eq. \eqref{eq:coded-mfp}) and MFP (see Eq. ({eq:amb-norm}-{eq:amb})) for locating a harmonic source ($f=150Hz$).   The SNR of the received data vector (see Eq. \ref{Noisy_Data}) was set to  $16$ dB.  For a fixed $M$ we aggregate performance statistics across 1000 simulations: 100 different source locations (chosen from $\mathcal{R}$ uniformly at random) and $10$ different draws of the $\bPhi_\omega$ for each location.  For each test simulation, the error between the true and estimated target location was recorded in units of the target ellipse radius (see Fig. \ref{fig:elong}). From the results of the $1000$ simulations, we calculated the  empirical distance tail probability $P_M(d)$-for a given number of random backpropagations $M$- as the fraction of results that produces a location estimate $\hat\vec{r}$ with $\|\hat\vec{r}-\vec{r_0}\|_\mathrm{e} > d$. As shown, we are able to estimate the target within the unit ellipse more than $99$\% of the time from only $M=6$ test vectors.  Notice that the cMFP actually outperforms the unnormalized version of the MFP (from \eqref{eq:amb} above) when $M\approx 6$.  This happens because the cMFP has an estimate of the normalizing factor in the denominator, as shown in \eqref{eq:coded-mfp}. The cMFP is really an estimate of the normalized MFP in \eqref{eq:amb-norm}, and indeed that formulation is what the cMFP approaches as the number of random backpropagations $M$ becomes equal to number of receivers $N$.

The cMFP was also tested in a variety of SNR for the single-frequency case. Fig.~\ref{fig:single}b shows the probability that the localization estimate is within the first ellipse (i.e. $d<1$) as a function of the number of random backpropagations $M$. In all cases, the failure probability asymptotically decreases exponentially in the number of random backpropagations. Finally, Fig.~\ref{fig:single}c shows the tail probability of distance error for a fixed number of random backpropagations $M=20$. As expected, the performance of cMFP gradually decrease as the SNR of the measurements is reduced from $16$dB to $0$dB, similarly to what occurs when using conventional MFP \cite{baggeroer1988matched}.

\begin{figure*}
	\centering
	\begin{tabular}{cc}
		\raisebox{0.5in}{\rotatebox{90}{\Large $\Pn_M \left\{\text{error} > d\right\}$}}
		\includegraphics[width=70.0mm]{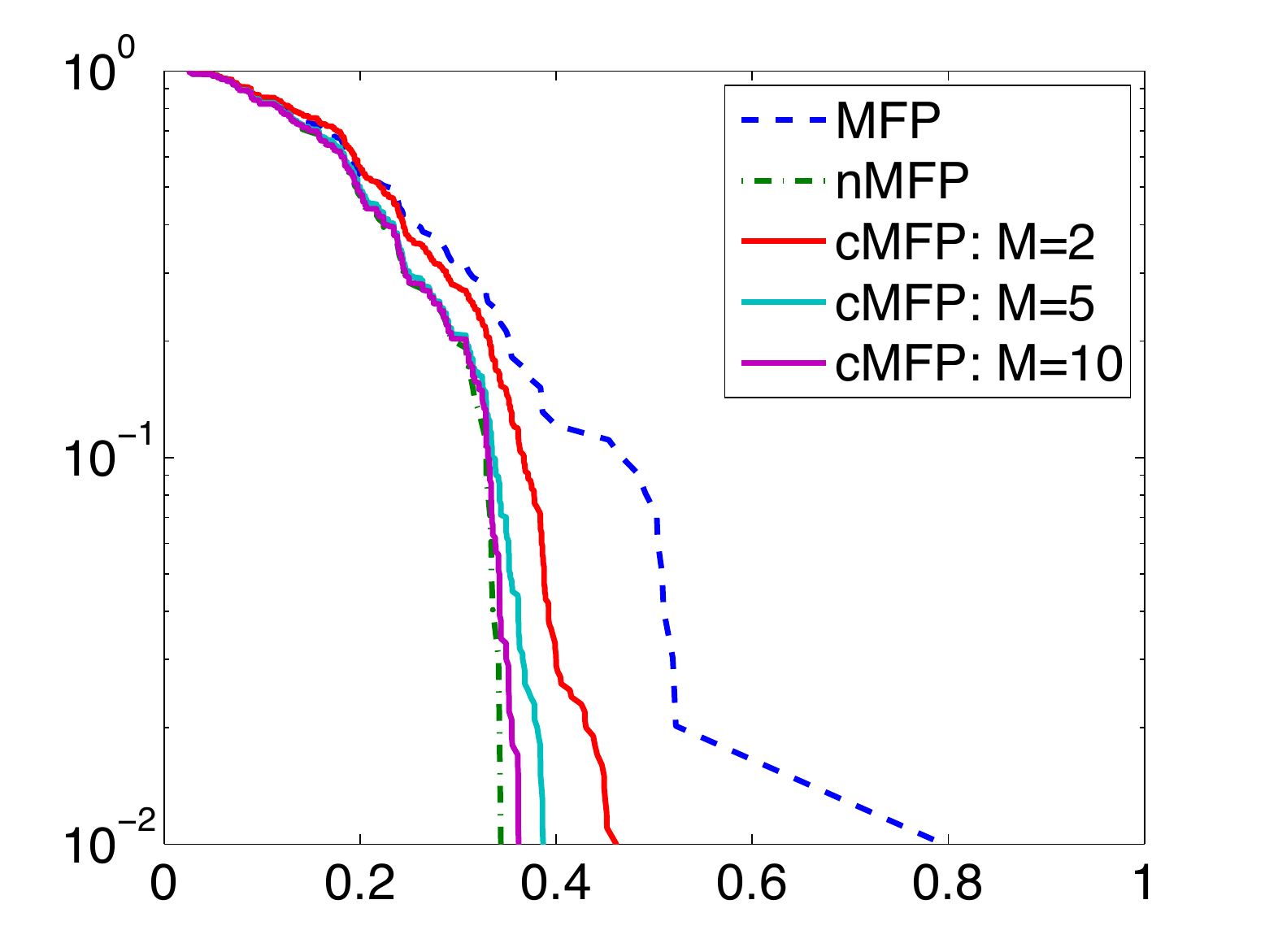} \\
		\hspace{.25in} Distance $d$ (in units of target ellipse radii) \\ (a) \\
		\raisebox{0.2in}{\rotatebox{90}{\Large $\Pn_M \left\{\text{error} > d=1\right\}$}}
		\includegraphics[width=70.0mm]{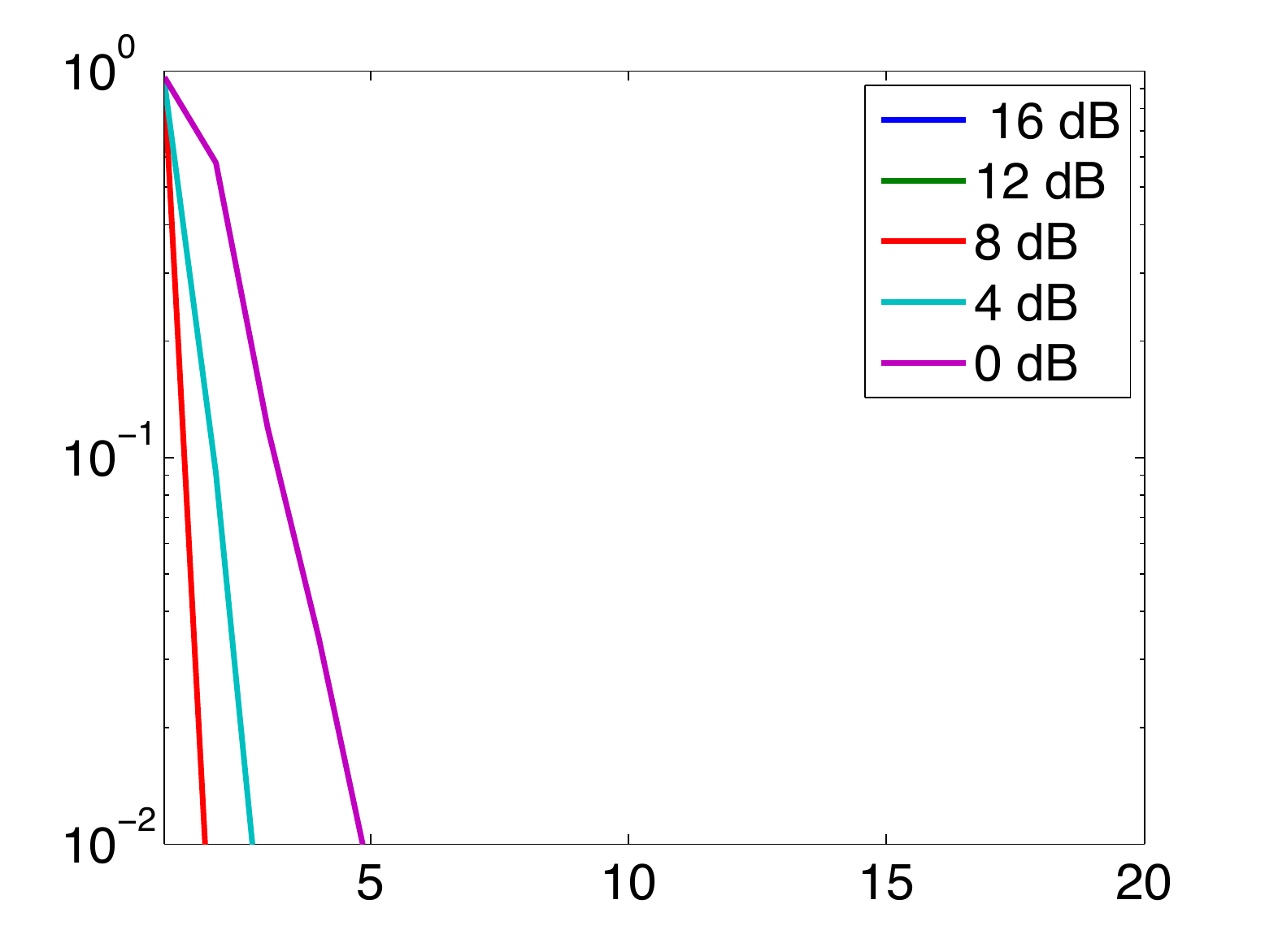} \\
		\hspace{.25in} Random Backpropagations: $M$ \\ (b) \\
		\raisebox{0.2in}{\rotatebox{90}{\Large $\Pn_{M=20} \left\{\text{error} > d\right\}$}}
		\includegraphics[width=70.0mm]{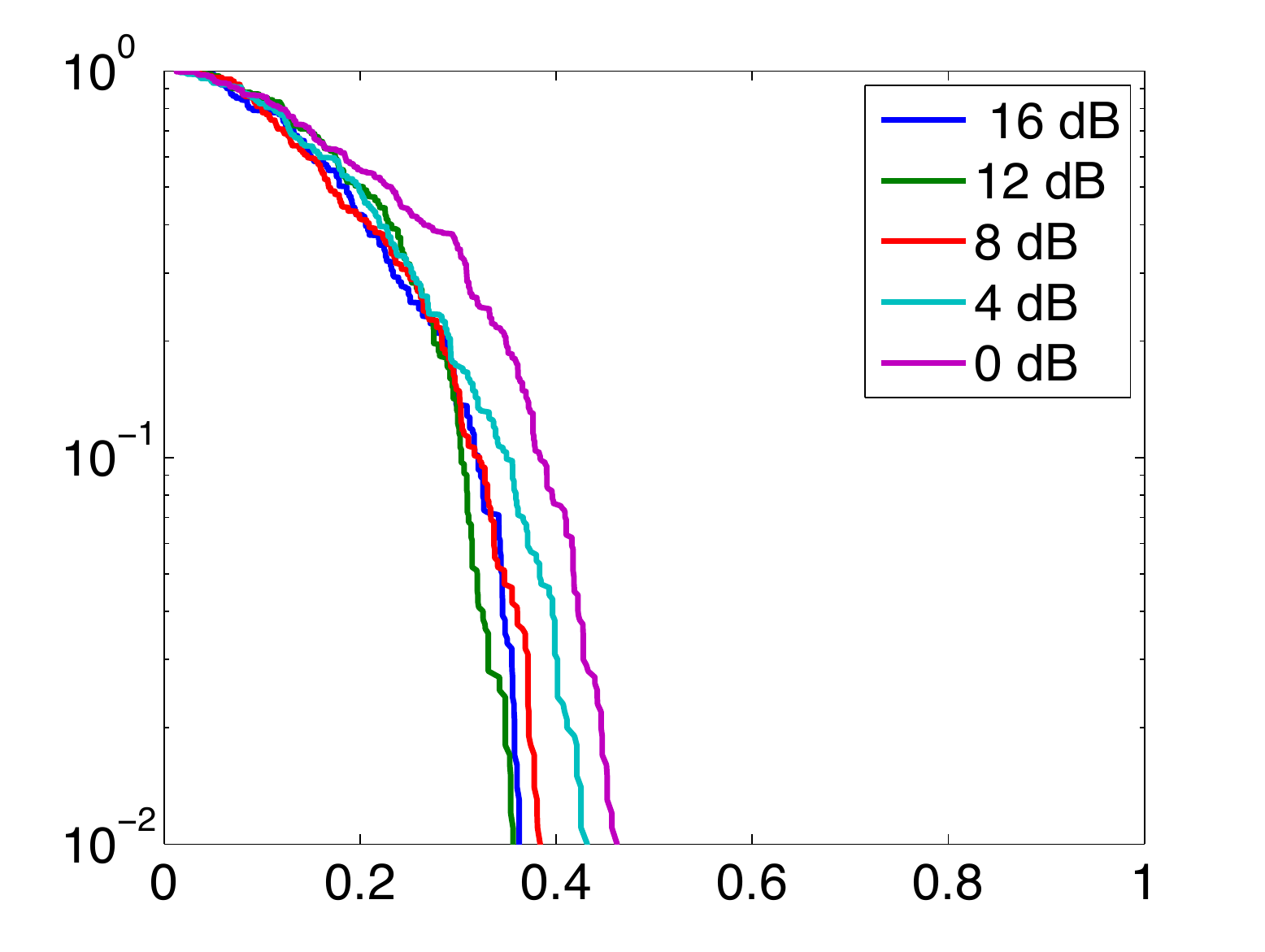} \\
		\hspace{.25in} Distance $d$ (in units of target ellipse radii) \\ (c)
	\end{tabular}
	\caption{\small\sl  
	Same as Fig. \ref{fig:single} but using instead the incoherent broadband cMFP formulation (see Eq. \eqref{eq:COHERENT_BROADBAND_CS_amb})
	}
	\label{fig:incoherent}
\end{figure*}

\begin{figure*}
	\centering
	\begin{tabular}{cc}
		\raisebox{0.5in}{\rotatebox{90}{\Large $\Pn_M \left\{\text{error} > d\right\}$}}
		\includegraphics[width=70.0mm]{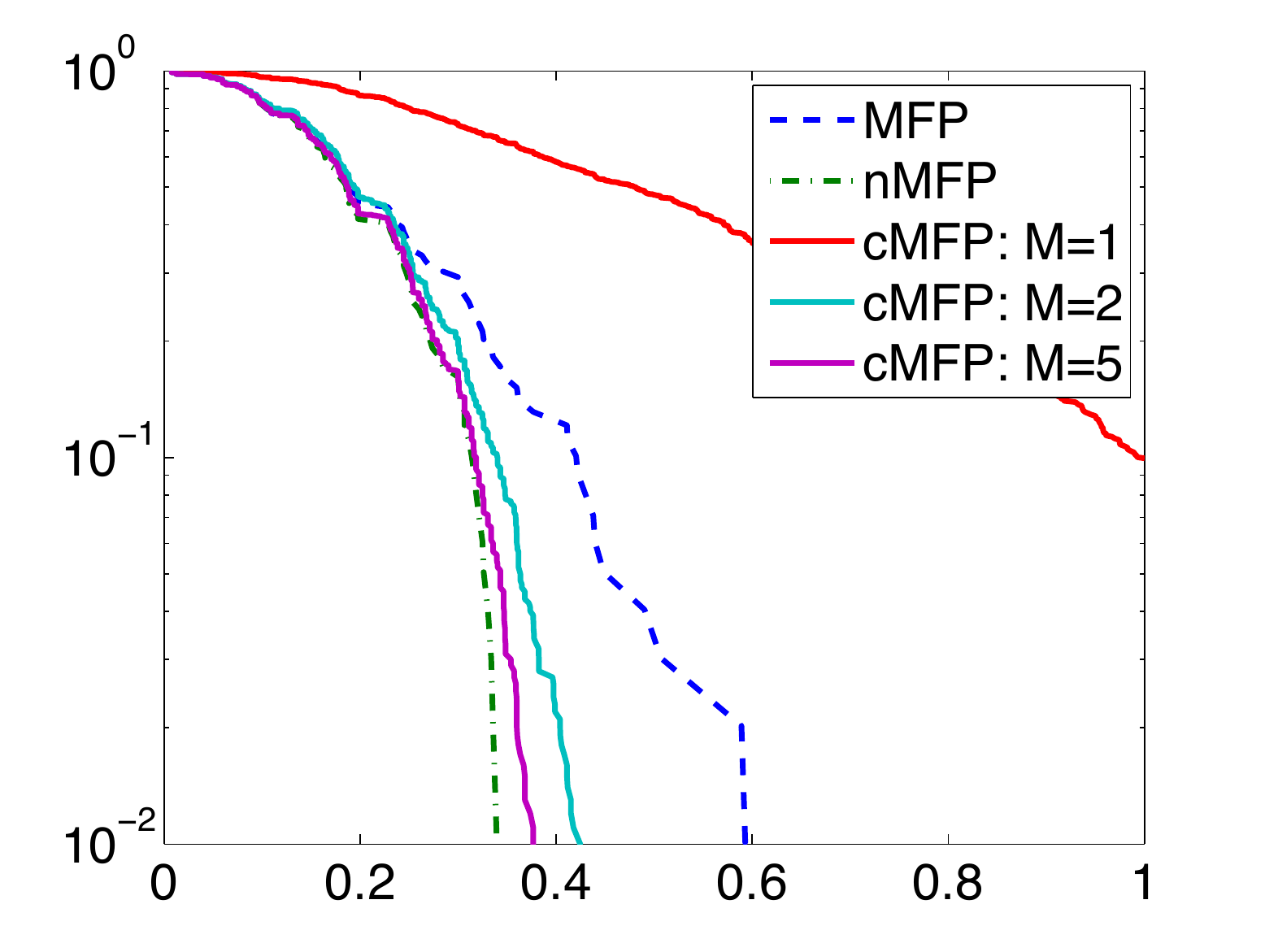} \\
		\hspace{.25in} Distance $d$ (in units of target ellipse radii) \\ (a) \\
		\raisebox{0.2in}{\rotatebox{90}{\Large $\Pn_M \left\{\text{error} > d=1\right\}$}}
		\includegraphics[width=70.0mm]{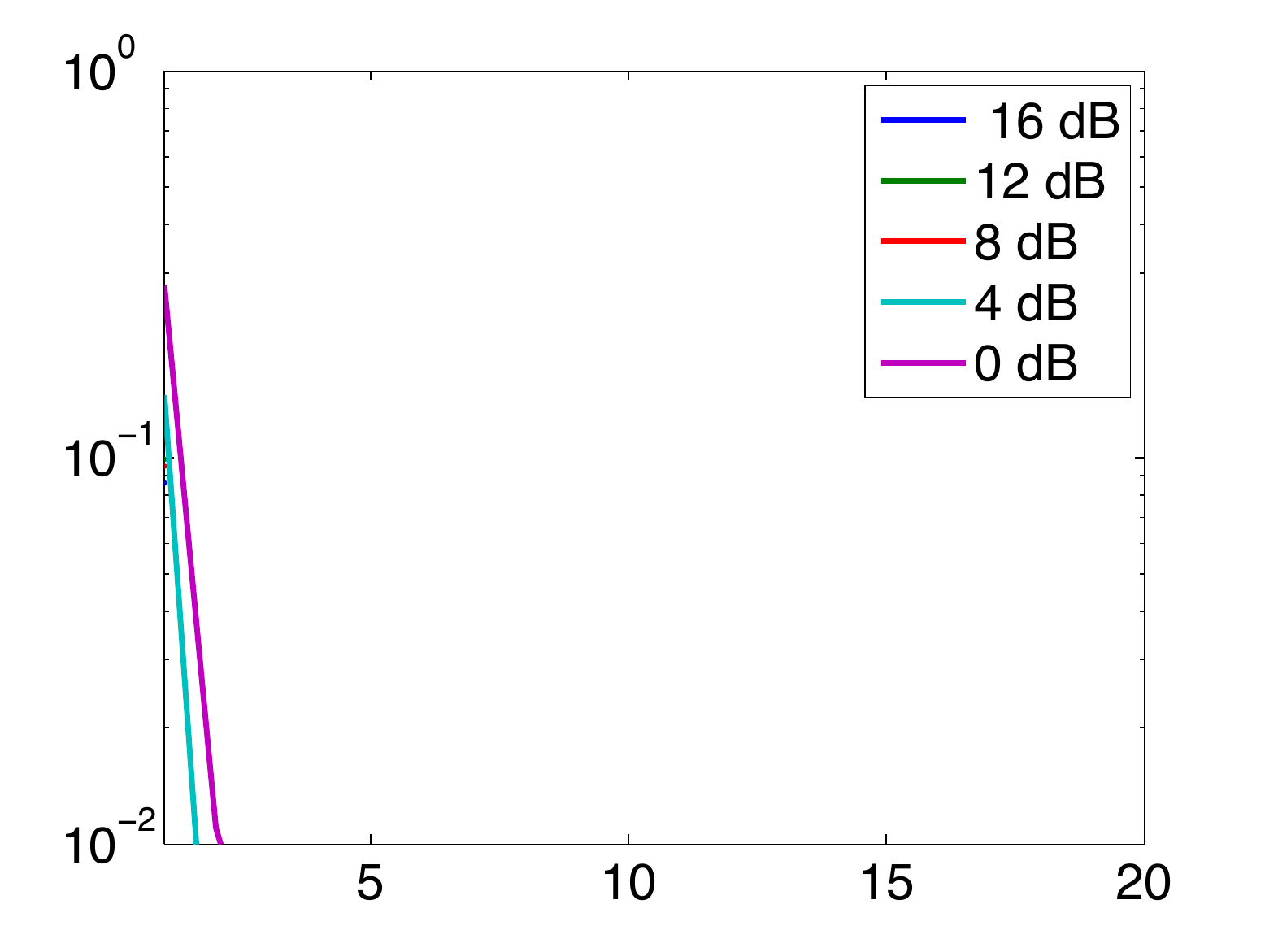} \\
		\hspace{.25in} Random Backpropagations: $M$ \\ (b) \\
		\raisebox{0.2in}{\rotatebox{90}{\Large $\Pn_{M=20} \left\{\text{error} > d\right\}$}}
		\includegraphics[width=70.0mm]{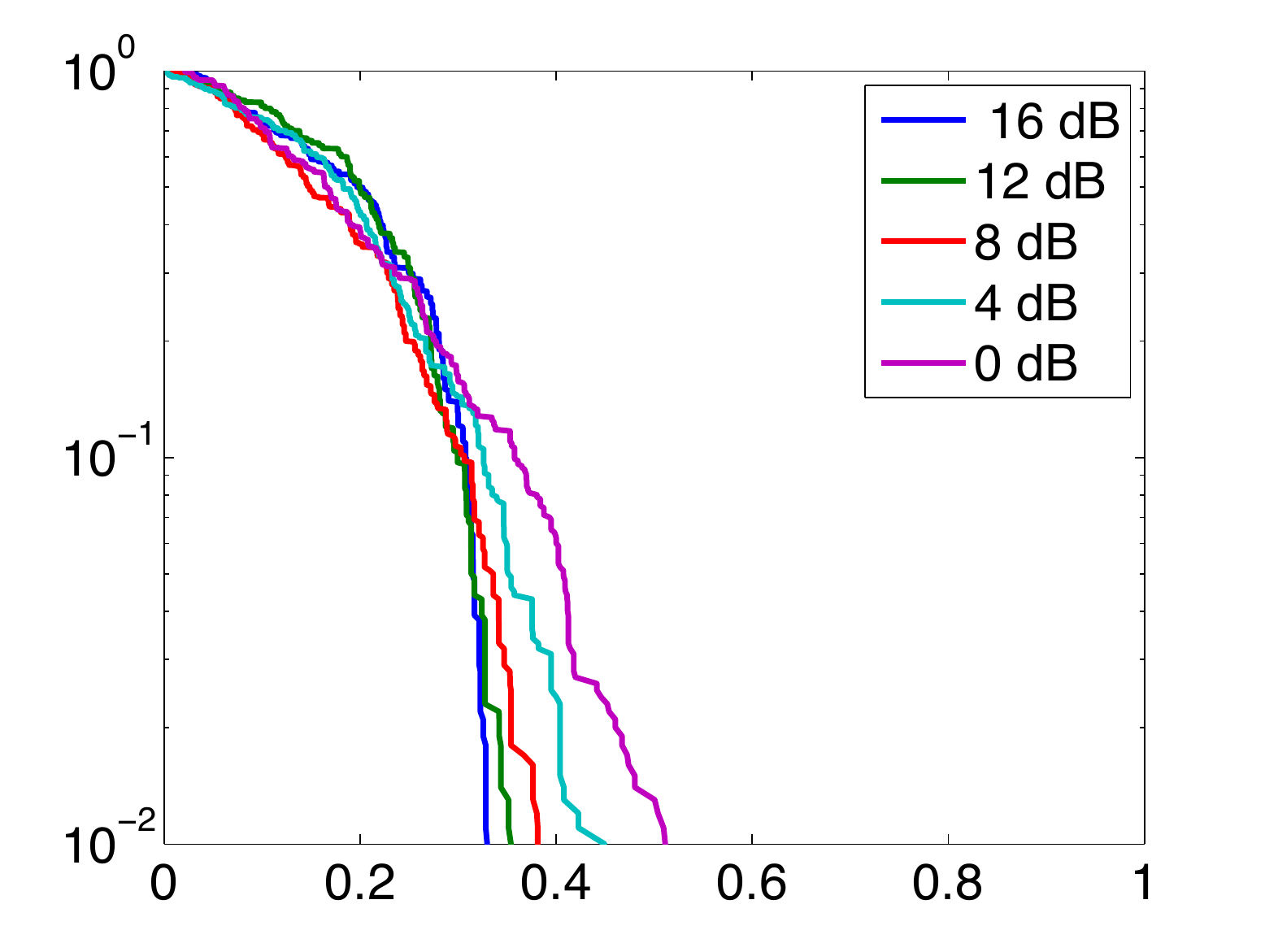} \\
		\hspace{.25in} Distance $d$ (in units of target ellipse radii) \\ (c)
	\end{tabular}
	\caption{\small\sl   
	Same as Fig. \ref{fig:single} but using instead the coherent broadband cMFP formulation (see Eq. \eqref{eq:COHERENT_BROADBAND_CS_amb})
}
\label{fig:coherent}
\end{figure*}

Fig.~\ref{fig:incoherent} and Fig.~\ref{fig:coherent} show a similar performance study for respectively the broadband incoherent cMFP (see Eq. \eqref{eq:INCOHERENT_BROADBAND_CS_amb}) or broadband coherent cMFP(see Eq. \eqref{eq:COHERENT_BROADBAND_CS_amb}) formulations, including the influence of the SNR of the measurements as well as the number  the number of random backpropagations $M$. Note that in Fig.~\ref{fig:coherent}, the  horizontal axis is normalized differently than in the other two cases due to the different spatial resolutions of the ambiguity surfaces (see Eq. \eqref{eq:elong}).  Our intentions are not to directly compare the performance of each of the three cMFP formulations (the coherent localization being always better as expected), but rather to show that in each case the selected cMFP formulation performs as well as the corresponding normalized MFP formulation and better the corresponding unnormalized MFP formulation.  This is especially true for the broadband coherent cMFP results as Fig.~\ref{fig:coherent}.a shows that with just $M=1$ measurement per frequency, we achieve an error within $3$ times what standard MFP gives us at least $90\%$ of the time, and with $M=2$, we fall within about $10\%$ distance error of what MFP gives us about $99\%$ of the time.

Furthermore, note that we do not show results for $M=1$ for the  broadband incoherent cMFP formulation (Fig.~\ref{fig:incoherent}.a) as in this case $\bPhi_kG_{\omega_k}(\vec{r})$ is a scalar for each $\vec{r}$, and \eqref{eq:INCOHERENT_BROADBAND_CS_amb} reduces to
\begin{align}
	\arg\max_{\vec{r}}\sum_{k=1}^K 
	\frac{|Y_{\omega_k}^H\bPhi_k^H\bPhi_k G_{\omega_k}(\vec{r})|^2}{\|\bPhi_k G_{\omega_k}(\vec{r})\|^2}
	&= \arg\max_{\vec{r}}\sum_{k=1}^K
	\frac{|\bPhi_kY_{\omega_k}|^2 |\bPhi_k G_{\omega_k}(\vec{r})|^2}
	{|\bPhi_k G_{\omega_k}(\vec{r})|^2} \\
	&= \arg\max_{\vec{r}}\sum_{k=1}^K |\bPhi_kY_{\omega_k}|^2.
\end{align}
This optimization problem is ill-defined, as the functional does not depend on $\vec{r}$.

\vspace{.2in}
\noindent
\subsection{Evolution of the main lobe to side lobe ratio of the cMFP ambiguity surface.}
\begin{figure*}
	\begin{tabular}{cc}
		\hspace{-5mm}
		\raisebox{0.2in}{\rotatebox{90}{main lobe/side lobe ratio}}
		\includegraphics[width=75.0mm]{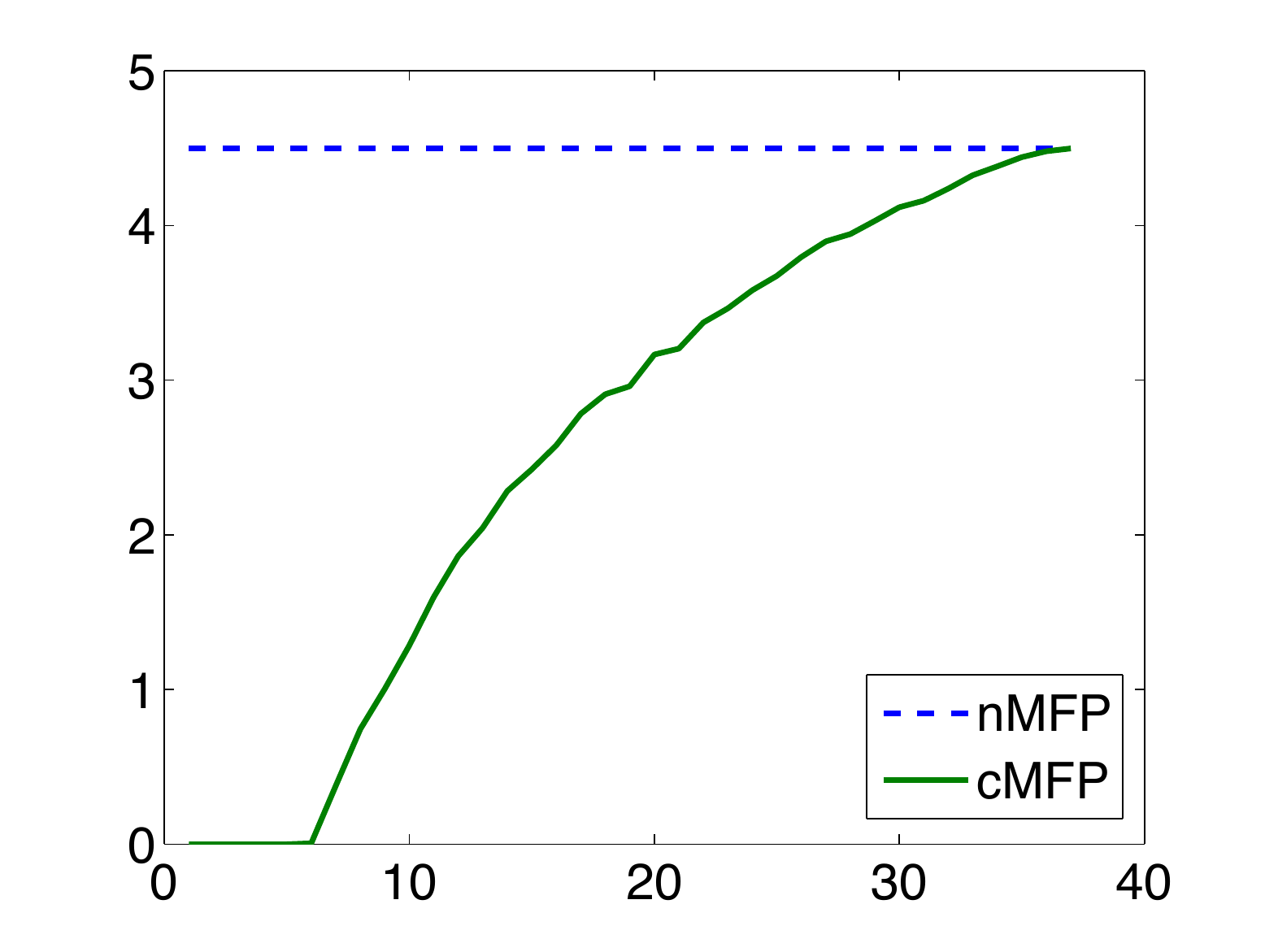} & \hspace{0in}
		\raisebox{0.2in}{\rotatebox{90}{main lobe/side lobe ratio}}
		\includegraphics[width=75.0mm]{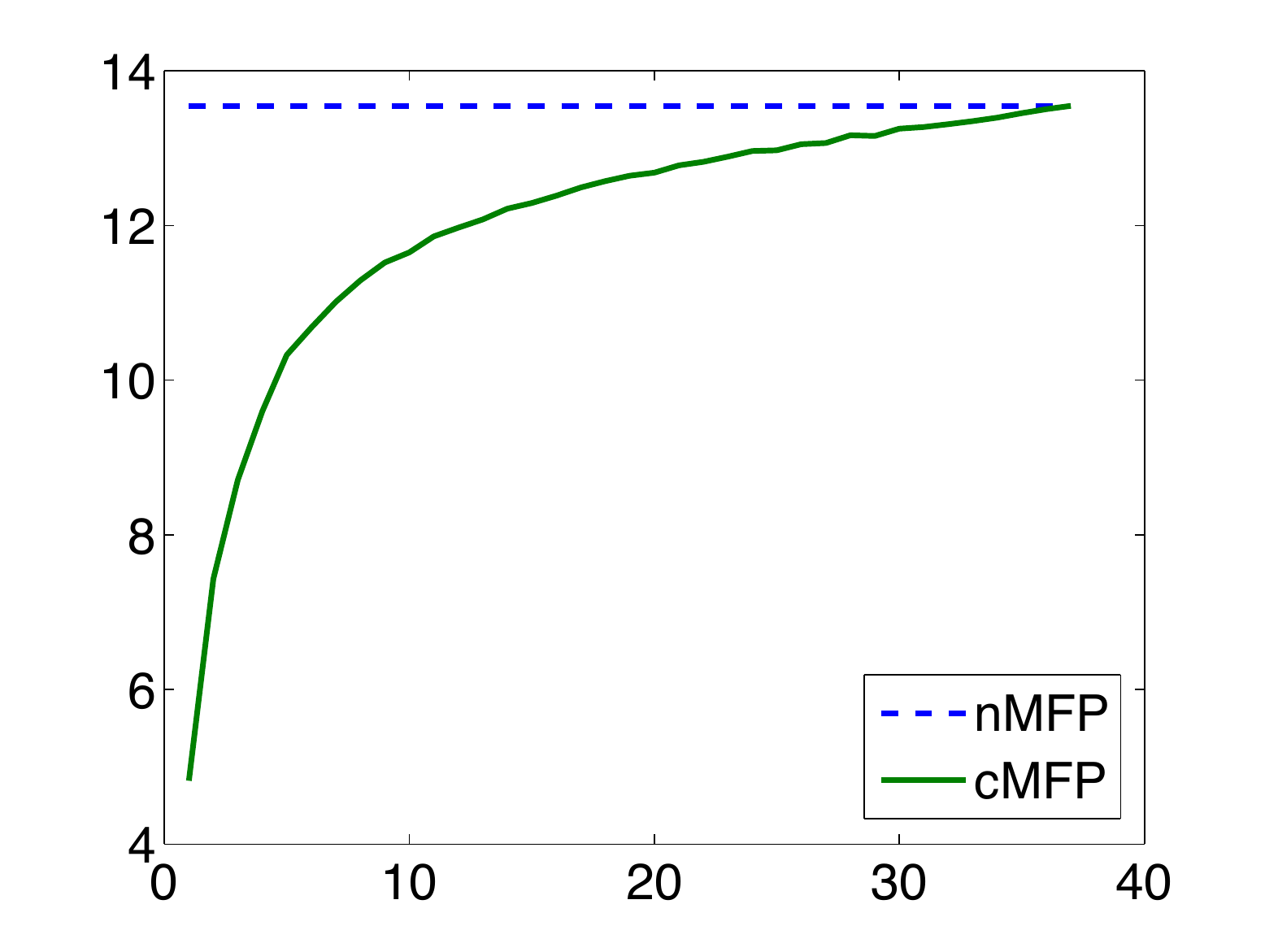} \\
		\hspace{.1in} Random Backpropagations: $M$ & \hspace{.4in} Random Backpropagations: $M$ \\
		(a) & \hspace{.4in} (b)
	\end{tabular}
	\caption{\small\sl  
	Evolution of the main lobe to side lobe ratio (in dB) of the estimated ambiguity surface (e.g. see Fig. \ref{fig:amb}) vs. number of random backpropagations $M$ using either (a) the single frequency cMFP formulation at $150$ Hz  or  (b) the broadband coherent MFP formulation (see Eq. \eqref{eq:COHERENT_BROADBAND_CS_amb}). Note that in each case the  main lobe to side lobe ratio of the ambiguity surface obtained with cMFP reaches the  main lobe to side lobe ratio value obtained using the corresponding nMFP formulation (dashed line) when $M=N=37$.
	}
\label{fig:lobe}
\end{figure*}

Fig.~\ref{fig:lobe} shows the logarithmic variations of the main lobe to side lobe ratio of the ambiguity surface obtained with the single frequency  and broadband coherent cMFP formulations for increasing number of random backpropagations $M$. In each case,  the displayed values represent the median value of the  main lobe to side lobe ratios obtained from $1000$ simulations for each value of $M$. Here the main lobe is defined as the maximum of the ambiguity surface $|h(\vect{r})|$ (obtained from the corresponding conventional MFP formulation, e.g. see Fig. \ref{fig:amb}a-b and Fig.~\ref{fig:cross}) over the region of interest $\mathcal{R}$, and the side lobe as the maximum of $|h(\vect{r})|$ over the search area $\mathcal{R}$ excluding an ellipse $E$  of the approximate size of the main lobe. We show the cross sections of the ambiguity function in Fig.~\ref{fig:cross} where we illustrate our choice of main lobe ellipse parameters that define our main lobe ellipse $E$. For the single frequency case, the ellipse has parameters $e^\mathrm{range} = 180$ meters and $e^\mathrm{depth} = 16$ meters (the broadband coherent case uses $e^\mathrm{range} = 72$ meters and $e^\mathrm{depth} = 16$ meters) as illustrated in Fig.~\ref{fig:cross}. The logarithmic value  of the main lobe to side lobe ratio is computed as:
\begin{equation}
	20 \log_{10}\left(\frac{\max_{\vect{r} \in \mathcal{R}} |h(\vect{r})|}{\max_{\vect{r} \in \mathcal{R} \setminus E} |h(\vect{r})|}\right),
\end{equation}
 Note that for small $M$ values, the cMFP side lobes may be significantly larger than their standard MFP counterparts.  The concentration inequality \eqref{eq:concentration} suggests that as $M$ gets larger, the side lobes dampen.  This behavior is observed in Fig.~\ref{fig:lobe}.  Note that since the $\bPhi$ matrix is an isometry when $M=N$, the side lobes in this case are exactly the same as for the standard MFP. 

\vspace{.2in}
\noindent

\subsection{Influence of  model mismatch on the cMFP performance}
\begin{figure*}
\centering
\begin{tabular}{c}
	\raisebox{20mm}{\rotatebox{90}{\bf Distance Error (m)}}
	\includegraphics[width=100.0mm]{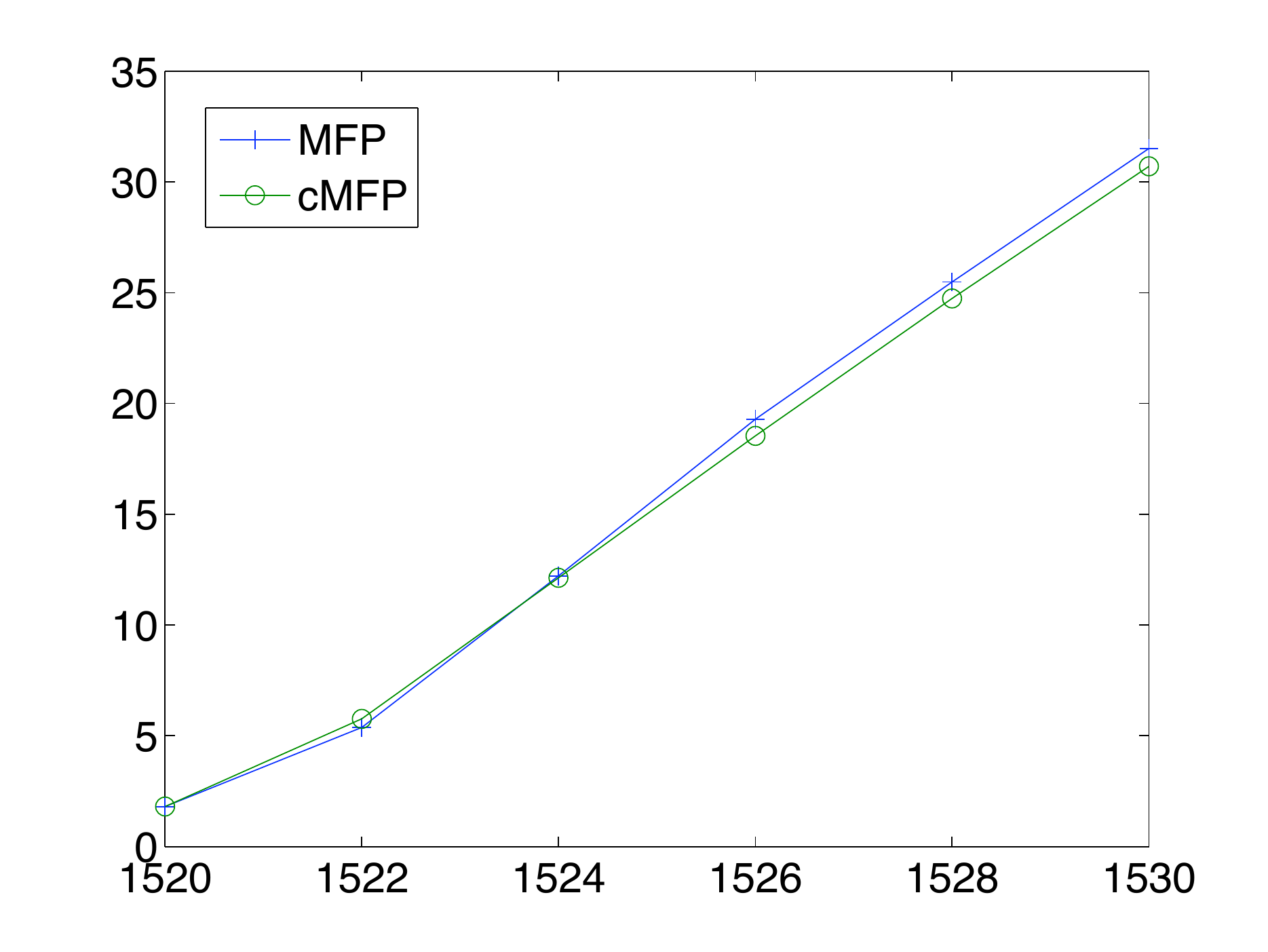} \\
	  { \bf Range (m)}
\end{tabular}
\caption{\small\sl 
Evolution of the localization error for  broadband coherent cMFP and corresponding conventional MFP a for increasing error of the modeled sound speed value. The correct sound speed value is $1520m/s$ here.    
 Notice that the localization errors obtained from cMFP (circle symbols) match  closely the  localization errors obtained obtained from standard MFP (cross symbols).
}
\label{fig:moderr}
\end{figure*}

Previous studies have shown extensively that one major liability of MFP is sensitivity to model mismatch which occurs when one has an incorrect model for the ocean waveguide (e.g. sound speed profile error) \cite{baggeroer1993overview}. Since MFP exploits the knowledge of the environment (via the Green's functions), its numerical accuracy must be sufficiently accurate, to ensure accurate source localization. Here we simply ensure that the localization accuracy of cMFP remains comparable to conventional MFP in the presence of error in the sound speed value. To do so, a set of received signals with a set SNR of $16$ dB were computed for a reference sound speed of $1520$ m/s. The broadband coherent cMFP -using $M=4$ random backpropagations per frequency (see \eqref{eq:COHERENT_BROADBAND_CS_amb}) and normalized MFP formulation (see \eqref{eq:nCoherent}) were then implemented using backpropagations in a simulated environment with different nominal values for the sounds speed (between $1520$ m/s and $1530$ m/s) than the reference value of $1520$ m/s. Fig.~\ref{fig:moderr} shows that the cMFP performs substantially the same as traditional MFP, for better or for worse. We show the average distance error in actual Euclidean distance (meters) as 
\begin{equation}
	\sqrt{(r_0^\mathrm{range}-r^\mathrm{range})^2+(r_0^\mathrm{depth}-r^\mathrm{depth})^2},
\label{eq:DistError}
\end{equation}
instead of ellipse distance. The small localization error occurring even without modeling error is due to the fact that the true source location did not coincide exactly with one the grid search location $\vec{r}$.

Note also that the range error tends to dominate for sufficiently large modeling error: the slope of the displayed error values is roughly $5000/1520$ (the nominal range divided by the nominal speed of sound) as we would expect because a $15$ m/s error in the speed of sound of $1520$ m/s causes a corresponding approximate $1\%$ distortion in the apparent range, or $50$ meters, i.e. $15/1520 \cdot 5000$.

\vspace{.2in}
\noindent
\subsection{Application of  cMFP for tracking a moving source.}

\begin{figure*}
\centering
\begin{tabular}{cc}
	\raisebox{20mm}{\rotatebox{90}{\bf Depth (m)}}
	\includegraphics[width=75.0mm]{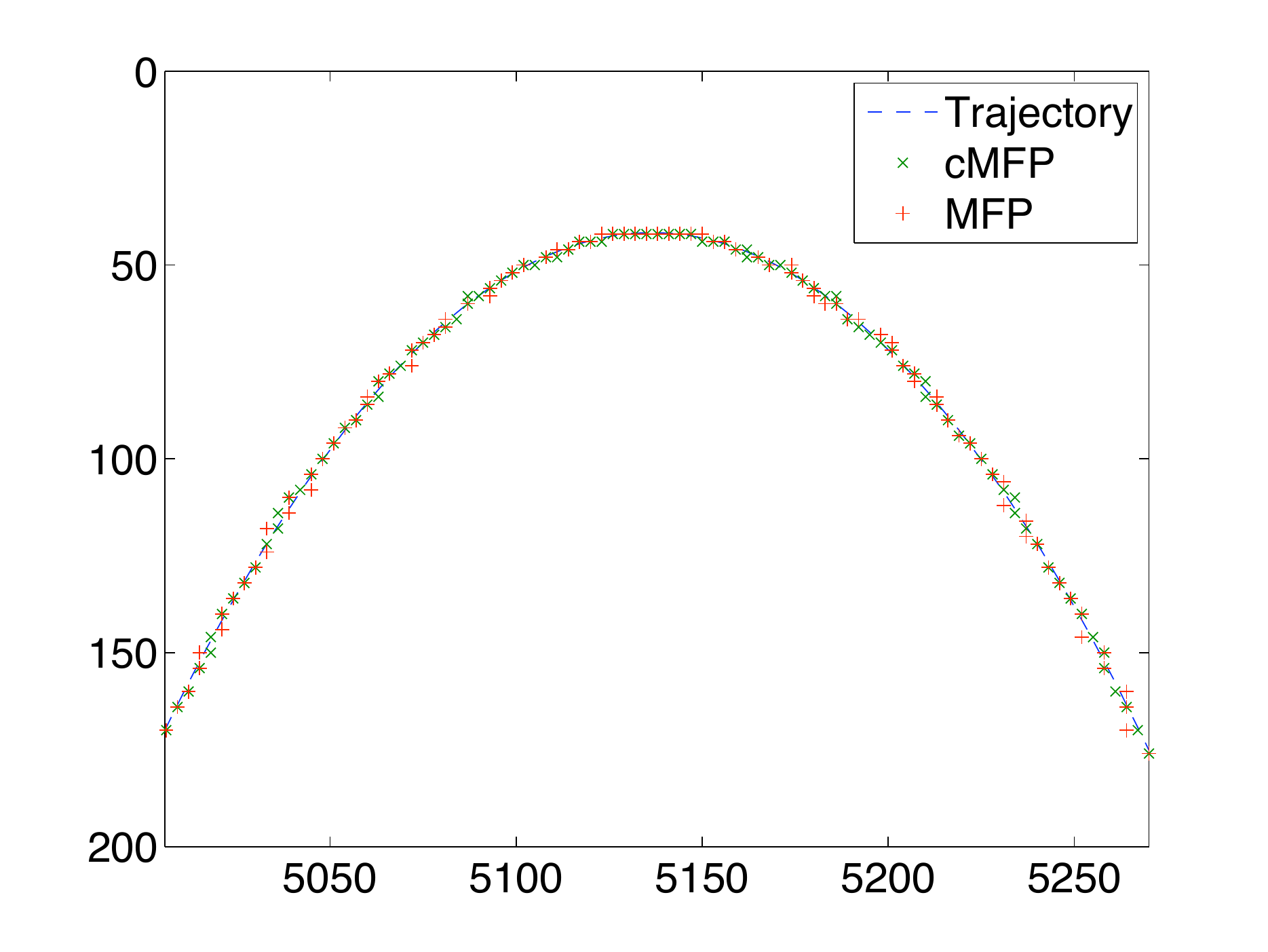} & \hspace{.2in}
	\includegraphics[width=75.0mm]{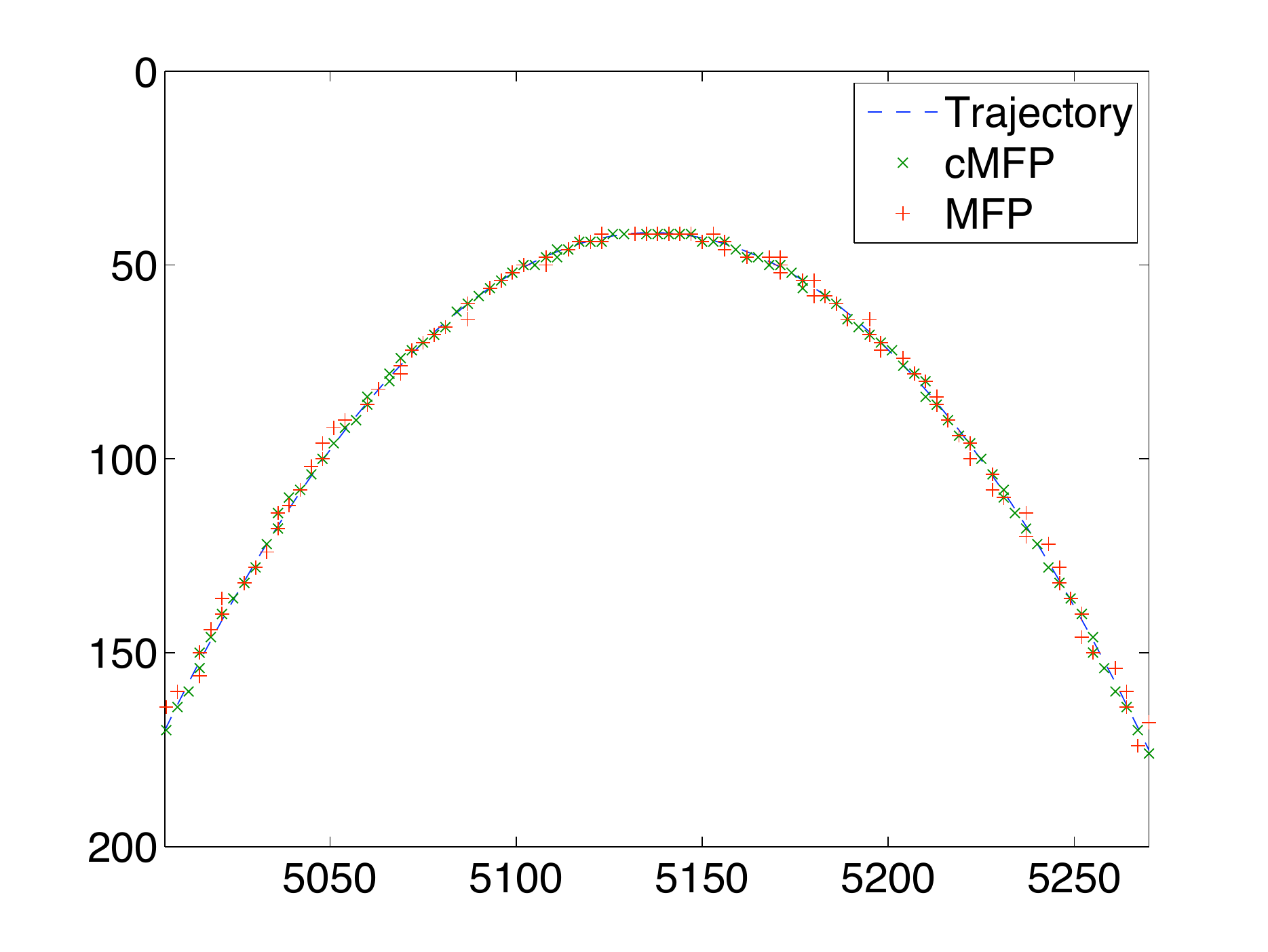} \\
	\hspace{.1in} { \bf Range (m)} & \hspace{.4in} { \bf Range (m)} \\
	(a) & \hspace{.4in} (b)
\end{tabular}
\caption{\small\sl 
Tracking of a source moving along a parabolic source trajectory (dashed line) using either coherent broadband cMFP, implemented with $M=2$ random backpropagations per frequency for the whole search grid, or using conventional broadband coherent MFP. For each  of the $100$ source positions, the SNR of the received signals at the vertical line array is constant and equal to (a) $16$dB or (b) $8$ dB.}
\label{fig:path}
\end{figure*}

The advantage of cMFP over conventional MFP for locating a moving source along a long track is illustrated here. Fig.~\ref{fig:path} displays the arbitrary path of a  source moving along a parabolic trajectory (dashed lines). For the sake of simplicity, the Doppler effect is not accounted:  this moving source scenario is simply simulated as  $100$ successive stationary sources located along the parabolic trajectory. For each positions, the SNR of the received signals at the vertical line array is constant and equal to  $16$ dB  (Fig.~\ref{fig:path}.a) or $8$ dB (Fig.~\ref{fig:path}.b). Conventional broadband coherent MFP is implemented by running $100$ successive backpropagations per frequency  over the search grid to estimate the source trajectory  (see crosshair symbols). On the other hand, broadband coherent cMFP is implemented using $M=2$ random backpropagations per frequency to estimate the same source trajectory  (see cross symbols).  The median value of the distance errors (computed from Eq. \eqref{eq:DistError}) between the estimated and actual source trajectory is $1$m when using both MFP and cMFP for a SNR of $16$dB. A slightly higher error of $1.6$m (resp. $1.1m$) for the cMFP (resp. MFP)  was found for a SNR of $8$dB. Overall, Fig.~\ref{fig:path} indicates that cMFP can potentially achieve comparable source tracking performance with a significantly reduced number of simulations.

\section{Extension to adaptive MFP}
\label{sec:adaptive}

\newcommand{\bQ}{\mathbf{Q}}

Several variants of the MFP algorithm have been proposed in the existing literature \cite{tolstoy2000applications,jensen1994computational} to enhance the robustness and performance of the basic Bartlett formulation presented above (see Eq. \eqref{eq:amb}). This can be especially beneficial in the presence of added coherent noise to the received data vector $Y_\omega$ (see Eq. \eqref{eq:themodel}). To do so, these higher resolution MFP algorithms are data adaptive, but typically have also a high resolution in their environmental knowledge requirements. A commonly used adaptive MFP formulation is the Minimum Variance Distortionless Response (MVDR) formulation. The MVDR formulation adaptively constructs a replica (or weighting) vector to yield a minimum mean square response to the recorded noise field along the receiver array while maintaining a constraint of unity processing gain for the incoming  signal vector $Y_\omega$ \cite[pg 540--552]{jensen1994computational}:
\begin{equation}
|	h^{MVDR}_\omega(\vec{r}) |^2=\left( \left(  G_\omega(\vec {r}) \right)^{H}  \bK^{-1} G_\omega(\vec {r}) \right)^{-1} .
\label{eq:amb_MVDR}
\end{equation}
where  $\bK$ is the $N \times N$ is the empirical correlation matrix from multiple realizations of the noisy received data vector $Y_\omega$:
\begin{equation}
\bK = \sum_{l=1}^L Y_{\omega,l} Y_{\omega,l}^H. 
\end{equation}
 The physical interpretation and performance analysis of the MVDR formulation (see Eq. \eqref{eq:amb_MVDR}) over the simple Bartlett formulation (see Eq. \eqref{eq:amb}) have been discussed extensively in the previous literature \cite{tolstoy2000applications,jensen1994computational} and thus will not be further repeated in this article.


The previous cMFP formulation can be readily extended to handled adaptive variants of the simple Bartlett MFP algorithm as discussed in Section III.C. For instance, using Eq. \eqref{eq:amb_MVDR} and by direct analogy to Eq. \eqref{eq:coded-mfp}, the magnitude square of the compressive MVDR ambiguity surface is:
\begin{equation}
|	\tilde{h}^{MVDR}_\omega(\vec{r}) |^2=\left( \left( \bPhi  G_\omega(\vec {r}) \right)^{H} \left( \bPhi \bK \bPhi^{H} \right)^{-1} \bPhi G_\omega(\vec {r}) \right)^{-1}.
\label{eq:amb_CS_MVDR}
\end{equation}
So once we have computed the $M$ tests measurements $\bPhi  G_\omega(\vec {r})$, they can be readily applied to either the compressive adaptive MFP formulation (see Eq. \eqref{eq:amb_CS_MVDR}) or the simple Bartlett formulation  (see Eq. \eqref{eq:coded-mfp}) to locate the unknown source.


\section{Conclusions}
\label{sec:conclusion}
We have shown here how dimension-reducing random projections can greatly reduce the computational cost involved with source localization via matched-field processing. When compared to the location of the maximum of the ambiguity surface obtained from conventional MFP using $N$ distributed receivers, the localization error achieved by cMFP scales down as square root of the number of random backpropagations $M$. The proposed cMFP formulation has also the added benefit to be able locate any source within the search grid area using only $M$ random backpropagations, while conventional MFP would require at least $N$ backpropagations to do the same. Thus cMFP provides an effective speedup factor of $N/M$ per frequency, which can be significant when a large number of receivers $N$ is available to locate a broadband source. Consequently this cMFP technique enables the ability to both broaden the search space and employ more sophisticated models of the Green's function, without introducing worries about sacrificing real-time performance


This compressive approach is not limited to source localization, and could be extended to a more general type of machine learning problem when matches are evaluated via inner products (or equivalently via Euclidean norms). This type of approach has the potential to substantially decrease computational complexity in these cases, while admitting a negligibly small probability of error.



\appendix

\section{Closest point on a line}
\label{app:closest-point}

For fixed vectors $U,V\in\C^n$, the following optimization program finds the closest point on the line spanned by $v$ to $u$,
\[
	\min_{\beta\in\C}~\|U - \beta V\|^2.
\]
The functional above attains its minimum value of
\[
	\|U\|^2 - \frac{|V^H U|^2}{\|V\|^2}
\]
when
\[
	\beta = \frac{V^H U}{\|V\|^2}.
\]	
This fact can be verified by differentiating $\|U-\beta V\|^2$ with respect to the real and imaginary parts of $\beta$, and solving for value of $\beta$ that makes them both equal to zero.

%
%



\bibliographystyle{plain}
\bibliography{arxiv_CMFP}


\end{document}